\documentclass[Journal]{IEEEtran}

\usepackage{amssymb,color,multirow,xcolor,colortbl}
\usepackage{amsmath}
\usepackage{graphicx}
\usepackage{epstopdf}
\usepackage{amsthm}
\usepackage{collcell, datatool}
\usepackage{subcaption}
\usepackage{cite}
\usepackage{pifont}
\usepackage{bbm}

\usepackage[normalem]{ulem}

\definecolor{lightgray}{gray}{0.9}

\begin{document}

\title{Modeling Cellular Networks with Full Duplex D2D Communication: A Stochastic Geometry Approach}
\author{
\IEEEauthorblockN{\large  Konpal Shaukat Ali, Hesham ElSawy, and Mohamed-Slim Alouini}

\thanks{The authors are with King Abdullah University of Science and Technology (KAUST), Computer, Electrical, and Mathematical Sciences and Engineering (CEMSE) Division,Thuwal, Makkah Province, Saudi Arabia. (Email: \{konpal.ali, hesham.elsawy, slim.alouini\}@kaust.edu.sa) 

This work was supported by the KAUST Sensor Research Initiative sponsored by the KAUST Office of Competitive Research. 
}}

\maketitle

\begin{abstract}
Full-duplex (FD) communication is optimistically promoted to double the spectral efficiency if sufficient self-interference cancellation {(SIC)} is achieved. However, this is not true when deploying FD-communication in a large-scale setup due to the induced mutual interference. Therefore, a large-scale study is necessary to draw legitimate conclusions about gains associated with FD-communication. This paper studies the FD operation for underlay device-to-device (D2D) communication sharing the uplink resources in cellular networks. We propose a disjoint fine-tuned selection criterion for the D2D and FD modes of operation. Then, we develop a tractable analytical paradigm, based on stochastic geometry, to calculate the outage probability and rate for cellular and D2D users. The results reveal that even in the case of perfect SIC, due to the increased interference injected to the network by FD-D2D communication, having all proximity UEs transmit in FD-D2D is not beneficial for the network. However, if the system parameters are carefully tuned, non-trivial network spectral-efficiency gains {({64\%} shown)} can be harvested. {We also investigate the effects of imperfect SIC and D2D-link distance distribution on the harvested FD gains.}
\end{abstract}

\IEEEpeerreviewmaketitle

\section{Introduction}
Improving spectral utilization has always been a core research focus in the field of wireless communication. A common technique to improve spectral utilization is to enable more aggressive spatial frequency reuse along with efficient interference coordination/cancellation. In this context, underlay device-to-device (D2D) communication is proposed to increase the spatial spectrum utilization in cellular networks. D2D communication allows proximity users equipments (UEs) to bypass the base station (BS) and communicate directly in a peer-to-peer fashion. While conventional cellular association prohibits intra-cell interference, D2D communication aggressively reuses the cellular spectrum over the spatial domain with no restriction over the cell boundaries. Consequently, D2D communication introduces a new type of interference between the cellular mode and D2D mode UEs, which is denoted as cross-mode\footnote{Cross-mode interference is used to denote both the cellular-to-D2D interference and the D2D-to-cellular interference} interference. Despite the increased interference level imposed by D2D communication, non-trivial gains can be harvested if efficient interference coordination between D2D and cellular links is adopted~\cite{3F,3G2,3H,3I,3J,3Kjrnl}. In addition to improving the spatial spectral utilization, D2D can potentially bring other performance gains for cellular networking, namely, lower power consumption, higher network capacity, and lower communication latency \cite{d2dSurvey1,d2dSurvey2,d2dSurvey3,d2dSurvey4,4A}.  

The performance gains offered by D2D communication are not sufficient to achieve the ambitious performance requirement defined for 5G networks \cite{5G_Jeff}. Therefore, it is believed that the foreseen 5G performance will be fulfilled by integrating several new technologies to the state of the art cellular system \cite{5G_Mag1}. 
In-band full-duplex (FD) communication is an appealing technology to integrate with D2D communication to further improve the spectral efficiency. FD communication exploits recent advances in transceiver design to mitigate the overwhelming self-interference (SI), and consequently, enables simultaneous transmission and reception on the same time/frequency resource blocks \cite{1sic1_fd5,2sic2_fd5}. Recent studies on a single-cell and single D2D link scenario have shown that FD D2D communication provides significant improvement in the spectral efficiency (up to $100\%$) over conventional half-duplex (HD) D2D if sufficient self-interference cancellation (SIC) is achieved \cite{FD1,FD2,FD3}. The studies in \cite{FD1,FD2,FD3} also emphasize the importance of cross-mode interference coordination in order to harvest the FD-D2D gains. However, the performance of FD-D2D communication in realistic large-scale setups has been overlooked. 

Since each transceiver can simultaneously transmit and receive on the same channel, FD-D2D communication activates two transmitters per D2D-link. Therefore, from a large-scale perspective where channels are reused over the spatial domain, FD-D2D communication can significantly increase the interference associated with D2D communication when compared to its HD counterpart. Note that cross-mode interference is already a performance limiting parameter for HD-D2D communication in cellular networks \cite{d2dOrig,d2dICC, 3F, 3I, 3J, 3Kjrnl}. Hence, it is hard to predict whether FD communication would improve or diminish the D2D gains due to the imposed interference. Therefore, studies for the FD-D2D effect on the aggregate interference in cellular networks are required. In this context, stochastic geometry provides a powerful mathematical tool that can be exploited to characterize the impact of interference associated with FD-D2D communication.

{Stochastic geometry has succeeded to provide a unified mathematical paradigm to model large-scale networks and characterize their operation~\cite{3B1,3B2,h_tut,3B3,3B5,3B6,di_renzo}. Using stochastic geometry, the operation of HD-D2D communication is studied in\cite{3F,3G2,3H,3I,3J,3Kjrnl,d2dSurvey1,d2dSurvey2,d2dSurvey3,d2dSurvey4,4A} and promising performance gains are reported. Also, several studies for FD communication in large-scale cellular networks are conducted via stochastic geometry \cite{FD4,FD5_2,FD6, AlAmmouri,Tsiky}, however, the FD communication is employed at the cellular link. When FD is employed at the cellular link, high rate improvement is observed for the downlink~\cite{FD4,FD5_2,FD6}. However, the authors in \cite{AlAmmouri,Tsiky} show that the downlink rate improvement may come at the expense of high degradation in the uplink rate due to the high disparity between the uplink and downlink transmit powers. The authors in \cite{FD5_2} report that FD communication offers rate improvement in both the forward and the reverse links when both have equivalent transmit powers. Although the model in \cite{FD5_2} is not for cellular networks, it motivates implementing FD communication to D2D links rather than to the cellular links due to the comparable UEs transmit powers. 
To validate the FD-D2D benefits, an explicit study for its operation in a large-scale setup is required.}

{In this paper, we develop a tractable analytical framework, based on stochastic geometry, for a single tier cellular network underlaid with D2D devices that share the cellular uplink resources and have FD communication enabled. The developed model accounts for a flexible D2D link distance distribution that captures different social interactions between the D2D devices. The UEs have limited transmit powers, employ truncated channel inversion power control, and follow a flexible D2D and FD/HD mode selection criterion. Based on the developed model, the FD-D2D enabled cellular network performance is assessed under perfect and imperfect self-interference assumptions. While imperfect SIC represents a practical operation scenario, perfect SIC has theoretical significance because it shows the explicit contribution of the FD-D2D communication to the aggregate interference level and reveals the subsequent effects on network performance. {Different from \cite{d2dOrig,d2dICC, 3F, 3I, 3J, 3Kjrnl} where the cellular network was overlaid with HD-D2D, a cellular network overlaid with FD-D2D is considered in this work. Additionally, different from \cite{FD5_2} where a FD ad-hoc network is considered, this work considers FD enabled in ad-hoc setting (D2D) overlaid with the cellular network.} The contributions and findings of the paper can be summarized as follows: }

\begin{itemize}

\item {The tradeoff, imposed by FD-D2D communication, between increasing the aggregate network interference and improving the spatial frequency reuse is mathematically modeled in terms of outage probability, defined as the probability that the SINR falls below a predefined threshold $\theta$, and the ergodic rate, defined by the seminal Shannon capacity formula.}

\item {We propose a flexible D2D and FD/HD mode selection and power control mechanism to balance the outage probability and spatial spectral efficiency tradeoff imposed by FD-D2D communication. The proposed mode selection and power control mechanisms are tailored to enable D2D communication, either in FD or HD modes, as long as a certain extent of  interference protection {(IP)} is enforced for cellular users. }

\item {The paper shows that enforcing FD-D2D communication may highly deteriorate the network performance due to the increased aggregate interference level in the network. On the other hand,  non-trivial gains can be harvested from the underlay FD-D2D communication with the proper design of the power control mechanism and D2D FD/HD mode selection criteria. For instance, the results show $64\%$ and $254\%$ spatial spectral efficiency gains harvested by the proposed FD-D2D communication when compared to the HD-D2D communication and conventional (i.e., D2D disabled) cellular network counterparts, respectively.  }


\item {The paper quantifies the gains that can be obtained by FD-D2D communication in terms of aggregate network throughput, per user throughput, and transmit power reduction. The paper also shows that there exist optimal values for the design parameters that maximize each of these gains. }



\item {From a mathematical perspective, an accurate approximation for the distance between the D2D-receiver and its closest BS is proposed, which is mandatory for modeling and designing the FD-D2D operation. }




\end{itemize}

The rest of the paper is organized as follows: Section \ref{SysMod} describes the system model. The statistics of the the link distances and UE classification are explained in Section \ref{LinkDis}. The analysis of transmit power statistics is covered in Section \ref{PowerAnalysis}. The assumptions made and SINR analysis are covered in Section \ref{SINRAnalysis}. Section \ref{Results} presents the results and Section \ref{Conc} concludes the paper.

\textit{\textbf{Notations:}} The mean of the RV $X$ is denoted by $\mathbb{E}[X]$. The probability of event $A$ is given by $\mathbb{P}(A)$. The ordinary hypergeometric function is denoted by ${}_2F_1(.,.;.;.)$. Also, we use $\gamma(m,n)=\int_0^n x^{m-1} e^{-x} dx$, $\Gamma(m,n)=\int_n^\infty x^{m-1} e^{-x} dx$, $\text{erf}(x)=\frac{2}{\sqrt{\pi}} \int_0^x  e^{-t^2} dt$, and $\text{erfc}(x)=1-\text{erf}(x)$. {A list of the symbols employed in this paper is given in Table \ref{Symbol_Table}.}

\section{System Model}\label{SysMod}
 
\subsection{Network Model}

\begin{table}[t]
\centering
\small
\caption{{\color{black} List of Symbols} }
\resizebox{0.4 \textwidth}{!}{\begin{tabular}{|c|c|}
\hline
\multirow{-2}{*}{}
&  \\
\multirow{-2}{*}{\textbf{Symbol}}&  \multirow{-2}{*}{\textbf{Definition}}  \\
 &  \\ \hline
    \multirow{2}{*}{\textbf{$\Psi$}}& \multirow{2}{*}{PPP to constitute cellular BSs}  \\
 &  \\ \hline
    \multirow{2}{*}{\textbf{$\Phi_c$}}& \multirow{2}{*}{PPP to constitute cellular UEs}  \\
 &  \\ \hline
    \multirow{2}{*}{\textbf{$\Phi_d$}}& \multirow{2}{*}{PPP to constitute D2D UEs}  \\
 &  \\ \hline
    \multirow{2}{*}{\textbf{$\omega$}}& \multirow{2}{*}{Control factor for D2D link distance distribution}  \\
 &  \\ \hline
    \multirow{2}{*}{\textbf{$\zeta$}}& \multirow{2}{*}{Fraction of residual SI}  \\
 &  \\ \hline
    \multirow{2}{*}{\textbf{$\lambda$}}& \multirow{2}{*}{Intensity of cellular BSs}  \\
 &  \\ \hline
    \multirow{2}{*}{\textbf{$\lambda_c$}}& \multirow{2}{*}{Intensity of cellular UEs}  \\
 &  \\ \hline
    \multirow{2}{*}{\textbf{$\lambda_d$}}& \multirow{2}{*}{Intensity of D2D UEs}  \\
 &  \\ \hline
\multirow{2}{*}{\textbf{$\rho_{min}$}}& \multirow{2}{*}{Receiver sensitivity}  \\
 &  \\ \hline
\multirow{2}{*}{\textbf{$\rho_c$}}& \multirow{2}{*}{Power control cutoff threshold at cellular receiver} \\
 &  \\ \hline
 \multirow{2}{*}{\textbf{$\rho_d$}}& \multirow{2}{*}{Power control cutoff threshold at f-D2D receiver}  \\
 &  \\ \hline
 \multirow{2}{*}{\textbf{$\rho_e$}}& \multirow{2}{*}{Power control cutoff threshold at r-D2D receiver}  \\
 &  \\ \hline
 \multirow{2}{*}{\textbf{$r_1$, $r_2$}}& \multirow{2}{*}{The ratios $\frac{\rho_c}{\rho_d}$, $\frac{\rho_d}{\rho_e} \big(=\frac{\rho_c}{\rho_e r_1}\big)$}  \\
 &  \\ \hline
 \multirow{2}{*}{\textbf{$r_2$}}& \multirow{2}{*}{The ratio $\frac{\rho_d}{\rho_e} \big(=\frac{\rho_c}{\rho_e r_1}\big)$}  \\
 &  \\ \hline
 \multirow{2}{*}{\textbf{$\bar{R}$}}& \multirow{2}{*}{Maximum D2D link distance}  \\
 &  \\ \hline
  \multirow{2}{*}{\textbf{$h$}}& \multirow{2}{*}{Small scale fading channel gain}  \\
 &  \\ \hline
  \multirow{2}{*}{\textbf{$\theta$}}& \multirow{2}{*}{Required SINR threshold}  \\
 &  \\ \hline
   \multirow{2}{*}{\textbf{$T_d$}}& \multirow{2}{*}{D2D bias factor}  \\
 &  \\ \hline
   \multirow{2}{*}{\textbf{$\mathcal{P}_d$}}& \multirow{2}{*}{Probability of f-D2D transmission}  \\
 &  \\ \hline
    \multirow{2}{*}{\textbf{$\mathcal{P}_e$}}& \multirow{2}{*}{Probability of r-D2D transmission}  \\
 &  \\ \hline
    \multirow{2}{*}{\textbf{$\mathcal{U}_d$}}& \multirow{2}{*}{Intensity of transmitting f-D2D UEs; $\mathcal{U}_d=\mathcal{P}_d \lambda_d$}  \\
 &  \\ \hline
    \multirow{2}{*}{\textbf{$\mathcal{U}_e$}}& \multirow{2}{*}{Intensity of transmitting r-D2D UEs; $\mathcal{U}_e=\mathcal{P}_e \lambda_d$}  \\
 &  \\ \hline
    \multirow{2}{*}{\textbf{$\mathcal{O}_p$}}& \multirow{2}{*}{Cellular truncation outage probability}  \\
 &  \\ \hline
    \multirow{2}{*}{\textbf{$\eta_c$}}& \multirow{2}{*}{Path-loss of cellular link}  \\
 &  \\ \hline
    \multirow{2}{*}{\textbf{$\eta_d$}}& \multirow{2}{*}{Path-loss of D2D link}  \\
 &  \\ \hline
    \multirow{2}{*}{\textbf{$P_c$}}& \multirow{2}{*}{Transmit power of generic cellular UE}  \\
 &  \\ \hline
    \multirow{2}{*}{\textbf{$P_d$}}& \multirow{2}{*}{Transmit power of generic f-D2D UE}  \\
 &  \\ \hline
    \multirow{2}{*}{\textbf{$P_e$}}& \multirow{2}{*}{Transmit power of generic r-D2D UE}  \\
 &  \\ \hline
    \multirow{2}{*}{\textbf{$P_u$}}& \multirow{2}{*}{Maximum transmit power of a UE}  \\
 &  \\ \hline
    \multirow{2}{*}{\textbf{$r_c$}}& \multirow{2}{*}{Generic cellular link distance}  \\
 &  \\ \hline
    \multirow{2}{*}{\textbf{$r_d$}}& \multirow{2}{*}{Generic D2D link distance}  \\
 &  \\ \hline
     \multirow{2}{*}{\textbf{$r_{c_2}$}}& \multirow{2}{*}{Distance from generic r-D2D UE to it's f-D2D UE's nearest BS}  \\
 &  \\ \hline
    \multirow{2}{*}{\textbf{$r_e$}}& \multirow{2}{*}{Distance from generic r-D2D UE to it's nearest BS ($r_e=r_{c_2}$ if}  \\
    \multirow{2}{*}{ }& \multirow{2}{*}{the same BS is nearest to both the f-D2D UE and r-D2D UE)}  \\
 &  \\ \hline
   \multirow{2}{*}{\textbf{$\sigma^2$}}& \multirow{2}{*}{Noise power}  \\
 &  \\ \hline
   \multirow{2}{*}{\textbf{$\chi$}}& \multirow{2}{*}{Mode of operation; $\chi \in \{c,d,e \}$ denotes the cellular,}  \\
   \multirow{2}{*}{ }& \multirow{2}{*}{f-D2D, and r-D2D modes of operation, respectively}  \\
 &  \\ \hline
   \multirow{2}{*}{\textbf{$\mathcal{I}_{\kappa \chi}$}}& \multirow{2}{*}{Interference from a transmitter in mode $\kappa$ to a receiver in mode $\chi$}  \\
 &  \\ \hline
   \multirow{2}{*}{\textbf{$\mathcal{R}_{\chi}$}}& \multirow{2}{*}{Link spectrum efficiency for a UE in mode $\chi$}  \\
 &  \\ \hline
  \multirow{2}{*}{\textbf{$\mathcal{T}_{avg}$}}& \multirow{2}{*}{Per-user rate}  \\
 &  \\ \hline
 \multirow{2}{*}{\textbf{$\mathcal{T}_n$}}& \multirow{2}{*}{Network throughput}  \\
 &  \\ \hline
\end{tabular}}
\label{Symbol_Table}
\end{table}
\normalsize

\begin{figure}
\begin{minipage}[htb]{0.98\linewidth}
\centering\includegraphics[scale=0.38]{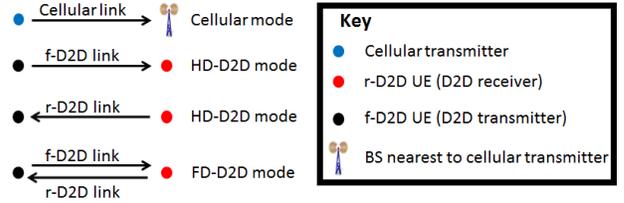}
\caption{Transmission Links and Modes}
\label{links}
\end{minipage}
\end{figure}
{We model a single-tier D2D-enabled cellular network, in which the D2D links are allowed to share the uplink cellular spectrum. The D2D UEs are equipped with FD transceivers and are allowed to operate in FD mode. Imperfect SIC is assumed such that $0\leq \zeta \leq 1$ fraction of the transmit power leaks back into the receiver chain of the FD transceiver.} We define the cellular link as the link from a UE to a BS, the forward-D2D (f-D2D) link as that from the D2D-transmitter to the D2D-receiver, and the reverse-D2D (r-D2D) link as that from the D2D-receiver to the D2D-transmitter. Hence, we refer to the D2D-transmitter that can transmit in f-D2D mode as the f-D2D UE, and the D2D-receiver that can transmit in the r-D2D mode as the r-D2D UE. {Note, we use the term D2D UEs to refer to both f-D2D and r-D2D UEs.} Consequently, the FD-D2D mode is active only if both the f-D2D link and r-D2D link are established between a D2D transmit-receive pair as shown in Fig. \ref{links}. As will be discussed later, we enable a flexible and disjoint mode selection scheme for the f-D2D and r-D2D links.

Independent Poisson Point Processes (PPPs) $\Psi$ and $\Phi_c$ are used to model the cellular BSs and the cellular UEs with intensities $\lambda$ and $\lambda_c$, respectively. We assume $\lambda_c \gg \lambda$ so that each BS always has a UE to serve. Cellular UEs associate to BSs based on the average radio signal strength (RSS), which reduces to the nearest BS association in single-tier networks. When multiple UEs associate to the same BS, they equally share its resources. The cellular network is overlaid by potential D2D transmitters modeled via an independent PPP $\Phi_d$ with intensity $\lambda_d$. Each D2D-transmitter (f-D2D UE) has a D2D-receiver (r-D2D UE) located within the D2D-proximity and can therefore bypass the BS and communicate in the D2D mode. \textcolor{black}{The D2D-proximity is defined as the region where the D2D-transmitter is able to invert path-loss and achieve at least a power of $\rho_{min}$ at its receiver while satisfying a maximum power constraint, where $\rho_{min}$ is the receiver-sensitivity.} Note, a D2D UE does not necessarily transmit in the D2D mode; it transmits in the D2D mode only if it satisfies the criteria required for D2D communication explained in the next subsection. D2D UEs that do not select the D2D mode are offloaded to out of band frequencies.\footnote{The offloading effect of D2D users to cellular mode and vice versa is studied in \cite{d2dOrig,d2dICC}. However, the offloading effect is not considered in this paper to avoid unnecessary complications to the analysis without providing additional insights.} The performance of such D2D nodes is out of the scope of this paper as they do not affect either the interference or the spectral efficiency within the band of interest.

{From the PPP assumption, the cellular link distance distribution, denoted by $r_c$, is given by $f_{r_c}(x)=2 \pi \lambda x e^{-\pi \lambda x^2}, \; x\geq 0$. There is no common agreement on the D2D link distance distribution in the literature as it may depend on the underlying application as well as the social interactions between the D2D UEs~\cite{3F, d2d_dist}. Therefore, we adopt the flexible distribution suggested in~\cite{d2d_dist}, which is given by }
\begin{equation}
f_{r_d}(x)=\frac{(2-\omega)x^{1-\omega}}{\bar{R}^{2-\omega}}, \; 0 \leq x \leq \bar{R}, 
\label{power_distance}
\end{equation}
{where $r_d$ is a random variable (RV) denoting the D2D link distance,  $\bar{R}=(\frac{P_u}{\rho_{min}})^{\frac{1}{\eta_d}}$ is the maximum transmission range of the D2D UE, and $ 0 \leq \omega < 2$ is a control factor for the distance distribution. Substituting $\omega =0$ in $f_{r_d}(\cdot)$ gives the no social interaction case where the D2D receiver is uniformly located in a circle with radius $\bar{R}$ around the D2D transmitter as in~\cite{d2dOrig}. Also, $\omega =1$ represents the case of equiprobable distances in the range of $[0,\bar{R}]$ as in~\cite{d2dICC}, and $ 1 < \omega < 2$ represents the case with high social interactions which gives higher weights to shorter D2D link distances. {It is worth noting that the distance from an f-D2D UE to its nearest BS is identical in distribution to $r_c$ and so we denote it by $r_c$ as well}. However, the distance from the r-D2D UE to its nearest BS {is denoted by $r_e$ and follows the distribution proposed in the next section.}

 A distance dependent power-law path-loss model is considered in which the signal power decays at the rate $r^{-\eta}$ with the distance $r$, where $\eta >2$ is the path-loss exponent. Since the D2D and cellular links may experience different propagation conditions, we discriminate between the path-loss exponents of the f-D2D link ($\eta_d$) and the cellular link ($\eta_c$). Assuming channel reciprocity, the r-D2D link has the same path-loss exponent as the f-D2D link. In addition to path-loss attenuation, transmitted signals experience Rayleigh fading with unit-mean exponential channel power gains. It is assumed that the channel gains are independent from the locations of the transmitters, receivers, and independent from one another.

It is assumed that all UEs have a unified maximum transmit power constraint of $P_u$. Due to the limited transmit power of the UEs, a truncated channel inversion power control is employed. Hence, only UEs that can compensate for the path-loss and maintain a predefined average power level at their receivers are allowed to transmit. The cutoff threshold for the power control of each of the communication modes is different; for link establishment we require the transmitters to maintain an average power of $\rho_{\chi}$ at their respective receivers, where $\chi \in \{c,d,e \}$ corresponds to $\{\text{cellular, f-D2D, r-D2D}\}$ modes. Such decoupled power control thresholds offer flexible network design and lead to an enhanced network performance. For the sake of simple presentation, we define $r_1=\frac{\rho_c}{\rho_d}$ and $r_2=\frac{\rho_d}{\rho_e}$. A cellular (f-D2D, r-D2D) connection can therefore be established if the power required to achieve $\rho_c$ ($\rho_d$, $\rho_e$) at the base station (r-D2D UE, f-D2D UE) does not exceed $P_u$, otherwise the transmitting UE goes into truncation outage. Due to the PPP assumption, the cellular-truncation outage probability can be expressed as $\mathcal{O}_p=e^{-\pi \lambda(\frac{P_u}{\rho_c})^{\frac{2}{\eta_c}}}\!\!\!\!\!$. 


Universal frequency reuse is assumed across the entire network with no intra-cell interference between cellular users. D2D links reuse the same uplink frequency with no restrictions on cell boundaries, but subject to the mode selection criterion described in the sequel. Without loss in generality, we analyze the system for one uplink channel.
\vspace{-2.5mm}

\subsection{Mode Selection}

{We consider a flexible mode selection criterion based on the bias factor $T_d$ to impose a tunable IP for the BSs. The IP is enforced via the following mode-selection inequalities $r_d^{\eta_d} \rho_d \leq T_d r_c^{\eta_c} \rho_c$ and $r_d^{\eta_d} \rho_e \leq T_d r_e^{\eta_c} \rho_c$, for the f-D2D and r-D2D UEs, respectively. In particular, the f-D2D does not operate in the D2D mode unless $r_d^{\eta_d} \rho_d \leq T_d r_c^{\eta_c} \rho_c$ is satisfied and the r-D2D does not operate in the D2D mode unless $r_d^{\eta_d} \rho_e \leq T_d r_e^{\eta_c} \rho_c$ is satisfied. These inequalities, denoted by IP conditions, ensure that a D2D link is not established unless the average interference power from the transmitting D2D UE (i.e. f-D2D or r-D2D) to its nearest BS is strictly less than $T_d \rho_c$, in which $T_d$ is a tunable design parameter to control the D2D contribution to the aggregate interference level. Consequently, $T_d$ controls the extent to which D2D is enabled in the network. Setting $T_d=0$ turns off D2D communication (both f-D2D and r-D2D) altogether and nullifies the D2D interference, while $T_d=\infty$ enforces D2D communication with no constraint on the D2D interference. Note that the D2D power control cutoff thresholds $\rho_d$ and $\rho_e$ can be also manipulated to encourage/discourage f-D2D and r-D2D link establishment, respectively, for a given $T_d$ without affecting the cellular IP (i.e., $T_d \rho_c$).} The employed mode selection scheme is summarized as follows:
\begin{itemize}
\item f-D2D UEs transmit in the f-D2D mode if they satisfy the IP-condition and {maximum transmit power constraint}. Otherwise, they go into truncation outage.\footnote{{Cellular truncation outage occurs due to unsatisfied power control cutoff threshold only. However, D2D truncation outage occurs due to either unsatisfied power control cutoff threshold and/or unsatisfied IP at the nearest BS.}}
\item r-D2D UEs transmit in the r-D2D mode if they satisfy the IP-condition and {maximum transmit power constraint}. Otherwise, they go into truncation outage.
\end{itemize}

\subsection{Methodology of Analysis}
We begin by analyzing the probability density functions (PDFs) of the link distances. This is followed by calculating the probabilities of transmitting in the f-D2D and r-D2D modes{, and the probability of a D2D pair being in FD}. The PDFs of the transmission powers in each mode of operation is then evaluated and the moments of the transmission powers are found. We characterize the SINR by its cumulative distribution function (CDF), which requires calculation of the Laplace transforms (LTs) of the interferences PDFs. We use the CDF of the SINR to infer link outage probability and spectral efficiency. To this end, we evaluate the FD-D2D enabled cellular network performance in terms of coverage, network spectral efficiency and power consumption. {For the sake of brevity, we define the \emph{network of interest} as a cellular network overlaid with FD-D2D that has BS intensity $\lambda$, D2D-UE intensity $\lambda_d$, D2D link distance distribution $f_{r_d}(\cdot)$, uniform f-D2D and r-D2D orientation, biasing factor $T_d$, and power control cutoff thresholds $\rho_c$, $\rho_d$, and $\rho_e$  for the cellular, f-D2D, and r-D2D modes, respectively.}

\section{On Link Distances and UE Classification}\label{LinkDis}
\subsection{Link Distance Analysis}

\begin{figure}
\begin{minipage}[htb]{0.98\linewidth}
\centering\includegraphics[scale=0.36]{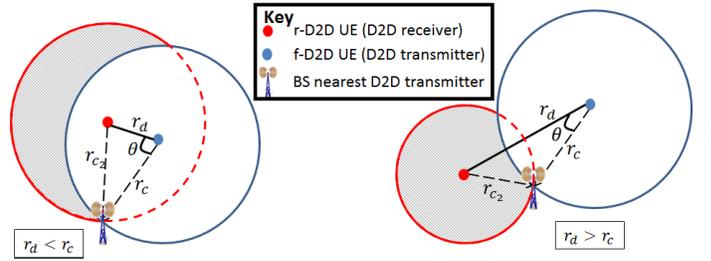}
\caption{The shaded crescent (with area $A$) which represents the region where a closer BS to the r-D2D may exist.}\label{crescents}
\end{minipage}
\end{figure}

{Based on the network realization and the relative positions of the f-D2D and r-D2D UEs, an f-D2D$-$r-D2D pair may or may not share the same nearest BS. Conditioning on the relative positions of the f-D2D UE, the r-D2D UE, and the nearest BS to the f-D2D UE, Fig.~\ref{crescents} shows two different instances of the shaded crescent {formed by the two disks, namely, the red disk centered at the r-D2D UE with radius $r_{c_2}$, and the blue disk centered at the f-D2D UE with radius $r_c$.} If a BS exists in the shaded crescent area, then the f-D2D and r-D2D UEs will not share a common nearest BS. Note that, by definition, a BS may only exist outside the circle of radius $r_c$ around the f-D2D UE, as the BS in Fig.~\ref{crescents} is the nearest BS to the f-D2D UE. Consequently, the region where the r-D2D UE may have a nearer BS is limited by the area of the shaded crescent shown in  Fig.~\ref{crescents}. The area of the shaded crescent depends on the relative values of the D2D link distance $r_d$, the distance between the f-D2D UE and its nearest BS $r_c$, and the distance between the r-D2D UE and the BS nearest to the f-D2D UE $r_{c_2}=\sqrt{r_c^2+r_d^2-2r_c r_d \cos \theta}$. Let $r_e$ be the distance between the r-D2D and its nearest BS. Then, $r_e=r_{c_2}$ if the shaded crescent contains no BS. Otherwise, $r_e < r_{c_2}$. Finding the distribution of $r_{c_2}$ is by itself a difficult problem because it is a function of three random variables, let alone the distribution of $r_e$ which is a function of $r_{c_2}$. Therefore, we propose a Rayleigh PDF approximation for the PDF of $r_e$ and verify its accuracy by simulations.}

The intuition behind our approximation is to use a Rayleigh distribution for $r_e$, which stems from the fact the the closest point from a 2-D PPP to any point in $\mathbb{R}^2$ follows the Rayleigh distribution. Exploiting the moment matching method, we only need the mean of $r_e$ for the Rayleigh distribution fitting. The following proposition formalizes the approximation of the distribution of $r_e$.

\textbf{\textit{Proposition 1:}} The distance between the r-D2D UE and its nearest BS (which may not be the same BS closest to its f-D2D UE) in the network of interest is accurately approximated by the following Rayleigh distribution:

\small
\begin{align}\label{eqn_fre}
 f_{r_e}(x)=\frac{\pi}{2 \mu_{r_e}^2} x \exp \Big( - \frac{\pi}{4 \mu_{r_e}^2} x^2 \Big) , \; x\geq 0,
\end{align}
\normalsize
where $\mu_{r_e}$ is an approximation of $\mathbb{E}[r_e]$ and is given by
\small
\begin{align}\label{mu_re_eqn}
&\mu_{r_e}=(1-\mathcal{P}_{r_e \neq r_{c_2}}) \mu_{r_{c_2}} + \nonumber \\
&\mathcal{P}_{r_e \neq r_{c_2}} \Big( \mathcal{P}_{r_c>r_d} \frac{\mu_{r_{c_2|r_c>r_d}} + \mathbb{E}[r_c|r_c>r_d]-\mathbb{E}[r_d|r_c>r_d]}{2} \nonumber \\
& + (1-\mathcal{P}_{r_c>r_d}) \frac{\mu_{r_{c_2}|r_c<r_d} + (\mathbb{E}[r_d|r_c<r_d]-\mathbb{E}[r_c|r_c<r_d])}{4} \Big).
\end{align}
\normalsize

\noindent In \eqref{mu_re_eqn}, $\mathcal{P}_{r_e \neq r_{c_2}}$ denotes the probability that the BS closest to the f-D2D UE is not the closest to the r-D2D UE and is given by $\mathcal{P}_{r_e \neq r_{c_2}}=1-e^{-\lambda A}$, where

\small
\begin{align}  \label{Area_A}
A= \mathbb{E}\Big[ r_{c_2}^2 (\phi - \frac{\sin(2\phi)}{2}) - r_c^2 (\bar{\theta} - \frac{\sin(2 \bar{\theta})}{2}) \Big]
\end{align}
 \normalsize
 
\noindent{is} the average area of the shaded crescents shown in Fig. \ref{crescents}, and $f_{\bar{\theta}}(\bar{\theta})=\frac{1}{\pi}, 0\leq \bar{\theta} \leq \pi$. The angle {\small $\phi=\pi-\arccos \Big(\frac{r_{c_2}^2+r_d^2-r_c^2}{2r_{c_2} r_d}\Big)$}. The probability that the f-D2D UE lies closer to the r-D2D UE than its nearest BS is {\small {$\mathcal{P}_{r_c>r_d}=\frac{2-\omega}{2\bar{R}^{2-\omega}} (\pi \lambda)^{\frac{\omega-2}{2}} [\Gamma(\frac{2-\omega}{2})-\Gamma(\frac{2-\omega}{2}, \pi \lambda \bar{R}^2)] $}}. 
 The mean of $r_{c_2}$ is approximated by {\small {$\mu_{r_{c_2}}= \sqrt{\frac{1}{4 \lambda} + \big(\frac{2-\omega}{3-\omega}\bar{R}\big)^2} $}}, 
 and the conditional expectations are given by
 \small
 {$\mu_{r_{c_2|r_c>r_d}}=\sqrt{(\mathbb{E}[r_c|r_c>r_d])^2 + (\mathbb{E}[r_d|r_c>r_d])^2}$} and\\
{ $\mu_{r_{c_2|r_c<r_d}}=\sqrt{(\mathbb{E}[r_c|r_c<r_d])^2 + (\mathbb{E}[r_d|r_c<r_d])^2}$}, \\
\normalsize
 where:\\
\small
{$\mathbb{E}[r_d|r_c>r_d]= \frac{1}{\sqrt{\pi \lambda}}\frac{\Gamma(\frac{3-\omega}{2})-\Gamma(\frac{3-\omega}{2}, \pi \lambda \bar{R}^2)}{\Gamma(\frac{2-\omega}{2})-\Gamma(\frac{2-\omega}{2}, \pi \lambda \bar{R}^2)} $} \\
{$\mathbb{E}[r_c|r_c>r_d]=\frac{4 (\pi \lambda)^{\frac{4-\omega}{2}}}{(2-\omega)[\Gamma(\frac{2-\omega}{2})-\Gamma(\frac{2-\omega}{2}, \pi \lambda \bar{R}^2)]}\Big[\frac{\Gamma(\frac{5-\omega}{2})-\Gamma(\frac{5-\omega}{2}, \pi \lambda \bar{R}^2)}{2 (\pi \lambda)^{\frac{5-\omega}{2}}} + \bar{R}^{2-\omega} \Big( \frac{\text{erfc}(\bar{R}\sqrt{\pi\lambda})}{4\pi \lambda^{1.5}} + \frac{\bar{R} e^{-\pi\lambda \bar{R}^2}}{2 \pi \lambda} \Big)\Big]  $} \\
{$\mathbb{E}[r_d|r_c<r_d]= \frac{2-\omega}{(1-\mathcal{P}_{r_c>r_d})}\Big(\frac{\bar{R}}{3-\omega}-\frac{(\pi \lambda)^{\frac{\omega-3}{2}}}{2\bar{R}^{2-\omega}}\big[\Gamma(\frac{3-\omega}{2})-\Gamma(\frac{3-\omega}{2}, \pi \lambda \bar{R}^2)\big]\Big) $} \\ 
{$\mathbb{E}[r_c|r_c<r_d]=  \frac{\frac{\text{erf}(\sqrt{\pi \lambda}\bar{R})}{2\sqrt{\lambda}}-\bar{R} e^{-\pi \lambda \bar{R}^2}}{(1-\mathcal{P}_{r_c>r_d})} - \frac{[\Gamma(\frac{5-\omega}{2})-\Gamma(\frac{5-\omega}{2}, \pi \lambda \bar{R}^2)]}{(\pi \lambda)^{\frac{3-\omega}{2}} \bar{R}^{2-\omega}(1-\mathcal{P}_{r_c>r_d})}  $}. \\ 
\normalsize

\textit{Proof:} See Appendix A.\qed\\

%
%
%
 Fig. \ref{verify_re} verifies the distribution of $r_e$ in Proposition 1 by plotting the CDF obtained from \eqref{eqn_fre} {for $\omega=1$}, against simulations. {Similar results are obtained for other values of $\omega$; which are not plotted for brevity}. Hereafter, we will use the notation $b=\frac{\pi}{4\mu_{r_e}^2}$ for the sake of simple exposition.


\begin{figure}
\begin{minipage}[htb]{0.9\linewidth}
\centering\includegraphics[scale=0.5]{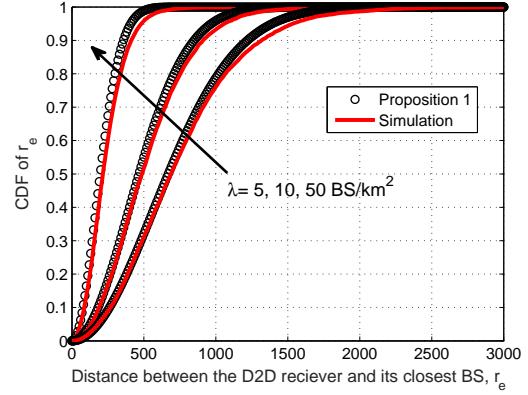}
\caption{CDF of the distance between the r-D2D UE and its closest BS, $r_e$, for {$\omega=1$ and} different BS intensities.}\label{verify_re}
\end{minipage}
\end{figure}

\subsection{UE Classification}

The probabilities that D2D transmitters and receivers select their respective modes of operation are given by the following:

\textbf{\textit{Lemma 1:}} The probability that an r-D2D UE in the \emph{network of interest} transmits in the r-D2D mode is given by

\small
{\begin{align*}
\mathcal{P}_e= \frac{(2-\omega)\eta_c}{2 \eta_d \bar{R}^{2-\omega}}   &\left( \frac{T_d\rho_c}{\rho_e b^{\frac{\eta_c}{2}}} \right)^{\frac{2-\omega}{\eta_d}} \gamma \left(\frac{(2-\omega)\eta_c}{2\eta_d}, b \Big(\frac{P_u}{T_d \rho_c}\Big)^{\frac{2}{\eta_c}} \right) .
\end{align*}}
\normalsize

\noindent The intensity of the r-D2D links (i.e., intensity of transmitting r-D2D UEs) is given by $\mathcal{U}_e=\lambda_d\mathcal{P}_e$.

\textit{Proof:} See Appendix B. \qed

\textbf{\textit{Lemma 2:}} The probability that an f-D2D UE in {the \emph{network of interest}} transmits in the f-D2D mode is given by

\small
{\begin{align*}
&\mathcal{P}_{d}= \frac{(2-\omega)\eta_c}{2\eta_d \bar{R}^{2-\omega}}\Big({\frac{T_d \rho_c}{(\pi \lambda)^{\frac{\eta_c}{2}}\rho_d}}\Big)^{\frac{2-\omega}{\eta_d}}\gamma\Big(\frac{(2-\omega)\eta_c}{2\eta_d},\pi \lambda \Big({\frac{P_u}{\rho_c T_d}}\Big)^{\frac{2}{\eta_c}}\Big). 
\end{align*}}
\normalsize

\noindent{The} intensity of the f-D2D links (i.e., intensity of transmitting f-D2D UEs) is given by $\mathcal{U}_d=\lambda_d\mathcal{P}_d$.

\textit{Proof:} See Appendix B. \qed

{
\textbf{\textit{Lemma 3:}} The probability that a D2D pair is in FD, i.e. both f-D2D and r-D2D UE are transmitting, in {the \emph{network of interest}} is given by 
\footnotesize
\begin{align*}
&\mathcal{P}_{FD}=\int_0^{\infty} \frac{f_{r_e}(g)/\bar{R}^{2-\omega}  }{ (1- \dot{q}) } \Bigg( \gamma \Big(\frac{(2-\omega)\eta_c}{2\eta_d}, \pi \lambda \Big(\frac{ \min(P_u, T_d g^{\eta_c} \rho_c)}{T_d \rho_c \rho_e/\rho_d}\Big)^{\frac{2}{\eta_c}}\Big)  \times \\ 
& \frac{(2-\omega)\eta_c}{2\eta_d} \Big(\frac{T_d \rho_c}{\rho_d (\pi \lambda)^{\frac{\eta_c}{2}}}\Big)^{\frac{2-\omega}{\eta_d}}  -  \Big(\frac{\min(P_u, T_d g^{\eta_c} \rho_c)}{\rho_e} \Big)^{\frac{2-\omega}{\eta_d}}\dot{q} \Bigg)  dg,
\end{align*}
\normalsize
where $\dot{q}=e^{-\pi \lambda (\frac{P_u}{\rho_c T_d})^{\frac{2}{\eta_c}} }$.}

\textit{Proof:} See Appendix C. \qed
\section{Transmit Power Analysis}\label{PowerAnalysis}

Due to the random network topology along with the employed truncated channel inversion power control, the transmit powers of the cellular, f-D2D, and r-D2D communication modes are all random variables. In this section, we characterize the PDF of the transmit powers of each mode as well as their moments.

\subsubsection{Forward-D2D Mode}
An f-D2D UE selects the f-D2D mode of operation if 1) it satisfies the maximum transmit power constraint, i.e. $r_d^{\eta_d}\rho_d<P_u$ 2) it satisfies IP to the cellular mode $r_d^{\eta_d}\rho_d<T_d r_c^{\eta_c}\rho_c$. The transmit power of a UE operating in the f-D2D mode can therefore be written as $P_d=r_d^{\eta_d}\rho_d$, with PDF given by the following Lemma.\\
\textit{\textbf{Lemma 4:}} In {the \emph{network of interest}} the PDF of the transmit-power of a UE operating in the f-D2D mode is given by,

\small
{\begin{align*}
f_{P_d}(x) &= \frac{2 x^{\frac{2-\omega}{\eta_d} - 1} e^{-\pi \lambda {(\frac{x}{T_d\rho_c})}^{\frac{2}{\eta_c}}} {(\pi \lambda)}^{\frac{(2-\omega)\eta_c}{2\eta_d}}}{\eta_c {(\rho_c T_d)}^{\frac{2-\omega}{\eta_d}} \gamma\Big(\frac{(2-\omega)\eta_c}{2\eta_d},\pi \lambda \Big({\frac{P_u}{\rho_c T_d}}\Big)^{\frac{2}{\eta_c}}\Big)},
\end{align*} }
\normalsize

\noindent{for} $0 \leq x \leq P_u$. The $\alpha^{th}$ moment of $P_d$ is given by,

\small
{\begin{align*}
\mathbb{E}[P_d^{\alpha}] &= \frac{(T_d \rho_c)^{\alpha} \gamma\Big(\frac{\alpha \eta_c}{2}+\frac{(2-\omega)\eta_c}{2\eta_d},\pi \lambda \Big({\frac{P_u}{\rho_c T_d}}\Big)^{\frac{2}{\eta_c}}\Big)}{{(\pi \lambda)}^{\frac{\alpha\eta_c}{2}}\gamma\Big(\frac{(2-\omega)\eta_c}{2\eta_d},\pi \lambda \Big({\frac{P_u}{\rho_c T_d}}\Big)^{\frac{2}{\eta_c}}\Big)}. 
\end{align*}}
\normalsize
\textit{Proof:} See Appendix D. \qed

\subsubsection{Reverse-D2D Mode}

An r-D2D UE selects the r-D2D mode of operation if 1) it satisfies the maximum transmit power constraint, i.e. $r_d^{\eta_d}\rho_e<P_u$, 2) it satisfies IP to the cellular mode of operation i.e. $r_d^{\eta_d}\rho_e<T_d r_e^{\eta_c}\rho_c$. The transmit power of an r-D2D UE operating in the r-D2D mode can therefore be written as $P_e=r_d^{\eta_d}\rho_e$, with PDF given by the following Lemma.

\textit{\textbf{Lemma 5:}}
In {the \emph{network of interest}}, the PDF of the transmit-power of a UE operating in the r-D2D mode is given by,

\small
{\begin{align*}
f_{P_e}(x) &=  \frac{2 b^{\frac{(2-\omega)\eta_c}{2 \eta_d}} x^{\frac{(2-\omega)}{\eta_d}-1} e^{-b(\frac{x}{T_d\rho_c})^{\frac{2}{\eta_c}}} }{\eta_c (T_d\rho_c)^{\frac{(2-\omega)}{\eta_d}}  \gamma \Big(\frac{(2-\omega)\eta_c}{2\eta_d}, b \Big(\frac{P_u}{T_d\rho_c }\Big)^{\frac{2}{\eta_c}} \Big)},
\end{align*}}
\normalsize

\noindent for $0\leq x \leq P_u$, and the $\alpha^{th}$ moment of $P_e$ is,

\small
{\begin{align*}
\mathbb{E}[{P_e}^{\alpha}] &=  \frac{(T_d \rho_c)^{\alpha} \gamma\Big( \frac{(2-\omega)\eta_c}{2\eta_d}+ \frac{\alpha \eta_c}{2} , b \Big( \frac{P_u}{T_d \rho_c} \Big)^{\frac{2}{\eta_c}} \Big) }{ b^{\alpha\frac{\eta_c}{2}} \gamma \Big(\frac{(2-\omega)\eta_c}{2\eta_d}, b \Big(\frac{P_u}{T_d\rho_c}\Big)^{\frac{2}{\eta_c}} \Big)}.
\end{align*}}
\normalsize
\textit{Proof:} See Appendix E. \qed

\subsubsection{Cellular Mode}

A cellular UE selects the cellular mode of operation when it is not in cellular truncation outage i.e. $r_c^{\eta_c}\rho_c<P_u$. The transmit power of the UEs operating in the cellular mode is written as $P_c=r_c^{\eta_c}\rho_c$, and the PDF is given by the following Lemma.

\textit{\textbf{Lemma 6:}} In {the \emph{network of interest}}, the PDF of the transmit-power of a UE operating in the cellular mode is,
\small
\begin{align*}
f_{P_c}(x) &= \frac{2 \pi \lambda x^{\frac{2}{\eta_c} - 1} e^{-\pi \lambda {(\frac{x}{\rho_c})}^{\frac{2}{\eta_c}}}}{\eta_c \rho_c^{\frac{2}{\eta_c}} \Big(1-e^{-\pi \lambda{(\frac{P_u}{\rho_c})}^{\frac{2}{\eta_c}}}\Big)}, \; \; \;  0 \leq x \leq P_u,
\end{align*}
\normalsize
\noindent and the $\alpha^{th}$ moment of the transmit power is given by,
\small
\begin{align*}
\mathbb{E}[P_c^{\alpha}] &= \frac{\rho_c^{\alpha} \gamma\Big( \frac{\alpha \eta_c}{2}+1, \pi \lambda \Big(\frac{P_u}{\rho_c} \Big)^{\frac{2}{\eta_c}} \Big)}{(\pi \lambda)^{\frac{\alpha \eta_c}{2}} \Big(1-e^{-\pi \lambda{(\frac{P_u}{\rho_c})}^{\frac{2}{\eta_c}}}\Big)}.
\end{align*}
\normalsize
\textit{Proof:} See Appendix F. \qed

\textcolor{black}{The intensity of the cellular UEs that are not in truncation is given by $(1-\mathcal{O}_p)\lambda_c$. Since only one UE is allowed to transmit per BS at a time on a given channel, the number of simultaneous transmitting cellular UEs on the same channel is limited by the number of BSs. Hence, the intensity of simultaneously active cellular UEs is limited by $\lambda$.}

\section{SINR Analysis}\label{SINRAnalysis}

{Let the Point Processes (PP) $\widetilde{\Phi}_c \subset {\Phi}_c$ and $\widetilde{\Phi}_d \subset {\Phi}_d$ denote the set of interfering cellular UEs and the set of interfering f-D2D UEs, respectively. Also, we define $\widetilde{\Phi}_e$ as the set of interfering r-D2D UEs. Although we have assumed ${\Phi}_c$ and ${\Phi}_d$ to be independent PPPs, neither $\widetilde{\Phi}_c$ nor $\widetilde{\Phi}_d$ is a PPP and both are mutually correlated due to their interactions (i.e., by scheduling and mode selection) with $\Psi$. Furthermore, $\widetilde{\Phi}_e$ is mutually correlated with $\widetilde{\Phi}_d$, and hence, is not a PPP. For tractability, we ignore the mutual correlations between $\widetilde{\Phi}_c$, $\widetilde{\Phi}_d$, and $\widetilde{\Phi}_e$, and assume that each of them constitutes an independent PPP. We formally state these approximations as follows:}\\
\textbf{Approximation 1:} The set of interfering cellular UEs ($\widetilde{\Phi}_c$) constitutes a PPP with with intensity $\lambda$, in which the transmit powers of the UEs are independent.\\
\textbf{Approximation 2:} The set of interfering f-D2D UEs ($\widetilde{\Phi}_d$) constitutes a PPP with intensity $\mathcal{U}_d$, in which the transmit powers of the UEs are independent.\\
\textbf{Approximation 3:} The set of interfering r-D2D UEs ($\widetilde{\Phi}_e$) constitutes a PPP with intensity $\mathcal{U}_e$, in which the transmit powers of the UEs are independent.\\
\textbf{Approximation 4:} The sets $\widetilde{\Phi}_c$, $\widetilde{\Phi}_d$, and $\widetilde{\Phi}_e$ are independent of one another.\\
{\textit{\textbf{Remark:}} It is worth mentioning that Approximations 1, 2, 3, and 4 only ignore the mutual correlations between interfering UEs. However, the correlation between the interfering UEs and the test-receiver is captured though the proper calculation of the IP boundaries. Similar approximations are done in \cite{d2dOrig,3B7,3F,JA1} for tractability, and are shown to be accurate. Such approximations maintain the model tractability and lead to simple yet accurate expressions for the distribution of the SINRs for each mode of operation. The accuracy of the aforementioned approximations and the distribution of $r_e$ in Proposition 1 are validated in Section \ref{Results} of this paper.}


We characterize the SINR by its CDF. For notational convenience we have defined the set $\chi \in \{c,d,e\}$ where $c$, $d$, and $e$ denote the cellular, the f-D2D, and r-D2D modes of operation, respectively. Hence, we can define a unified SINR expression for all modes of operation as

\begin{align*}
\text{SINR}_{\chi}=\frac{\rho_{\chi} h_0}{\sigma^2 + \mathcal{I}_{c\chi}+ \mathcal{I}_{d\chi}+ \mathcal{I}_{e\chi} {+ \zeta P_{\chi} \mathbbm{1}_{FD}}},
\end{align*}

\noindent{where} the noise power is denoted by $\sigma^2$, $\mathcal{I}_{\kappa\chi}$ is the interference from UEs transmitting in mode $\kappa$ ($\in \{c,d,e\}$) to the receiver of the UE transmitting in mode $\chi$, {and $\mathbbm{1}_{FD}$ is the event that both the f-D2D and r-D2D UEs are active i.e. the FD-D2D mode is active. For $\chi=c$, $\mathbbm{1}_{FD}=0$; for $\chi\in \{d,e\}$, $\mathbbm{1}_{FD}$ is 1 with probability $\mathcal{P}_{FD}$ and is 0 otherwise.} The interference $\mathcal{I}_{\kappa\chi}=\sum_{u_i \in \widetilde{\Phi}_{\kappa}} P_{\kappa_i} h_i ||y-u_i||^{-\eta_{\chi}}$, where $y$ and $u_i$ denote the positions of the test receiver and the $i^{th}$ interferer, respectively, $P_{\kappa_i}$ denotes the transmit power of the $i^{th}$ interferer, and $h_i$ denotes the channel between the $i^{th}$ interferer and receiver. The SINR outage is evaluated as:

\small
\begin{align}
& \mathbb{P}(\text{SINR}_{\chi} \leq \theta)= \mathbb{P}( h_0 \leq \frac{\theta}{\rho_{\chi}} (\sigma^2 + \mathcal{I}_{c \chi} + \mathcal{I}_{d \chi}  + \mathcal{I}_{e \chi} {+ \zeta P_{\chi} \mathbbm{1}_{FD}})) \nonumber \\
&= 1- e^{- \frac{\theta}{\rho_{\chi} } (\sigma^2 + \mathcal{I}_{c \chi} + \mathcal{I}_{d \chi} + \mathcal{I}_{e \chi}{+ \zeta P_{\chi} \mathbbm{1}_{FD}})} \nonumber \\
&=  1- e^{- \frac{\theta}{\rho_{\chi} } \sigma^2}  { \mathcal{L}_{P_{\chi}} \Big( \frac{\theta \zeta \mathbbm{1}_{FD}}{\rho_{\chi} } \Big)} \underset{\kappa \in \{c,d,e\}}\prod \mathcal{L}_{\mathcal{I}_{\kappa \chi}} \Big( \frac{\theta}{\rho_{\chi} } \Big).\label{BlahSINR_eqnGen}
\end{align}
\normalsize

\noindent{where} the second equality follows from the exponential distribution of $h_0$, and $\mathcal{L}_X(s)$ denotes the LT of the PDF of the RV $X$ evaluated at $s$. It is worth noting that at the event $\mathbbm{1}_{FD} = 0$, the LT $\mathcal{L}_{P_{\chi}}(0) =1$. {In particular, when the imperfect SIC scenario is considered, the SINR outage for the f-D2D and r-D2D UEs ($\chi \in \{d,e \}$) is calculated as $\frac{\mathcal{P}_{FD}}{\mathcal{P}_{\chi}} \mathbb{P}(\text{SINR}_{\chi} \leq \theta | \mathbbm{1}_{FD}=1)+ (1-\frac{\mathcal{P}_{FD}}{\mathcal{P}_{\chi}}) \mathbb{P}(\text{SINR}_{\chi} \leq \theta | \mathbbm{1}_{FD}=0)$. The weights account for the fraction of the D2D UEs transmitting in FD and HD, respectively. For the perfect SIC scenario, \eqref{BlahSINR_eqnGen} can be used directly (with $\mathcal{L}_{P_{\chi}}(0) =1$).} The LTs for the aggregate interferences are given by the following lemma. \\
\textbf{\textit{Lemma 7}:} For {the \emph{network of interest}}, the LTs of the interferences PDFs are:

\small
\begin{align*}
& \mathcal{L}_{\mathcal{I}_{ec}}(s)=\exp \Bigg(-\frac{ s \mathbb{E} \Big[ P_e^{\frac{2}{\eta_c}} \Big]{}_2F_1 \Big(1,\frac{\eta_c-2}{\eta_c};\frac{2\eta_c-2}{\eta_c};-s \rho_c T_d \Big)  }{ (2\pi \mathcal{U}_e)^{-1} (\rho_c T_d)^{\frac{2}{\eta_c}-1} (\eta_c-2) } \Bigg)\\
& \mathcal{L}_{\mathcal{I}_{dc}}(s)=\exp \Bigg(-\frac{ s \mathbb{E} \Big[ P_d^{\frac{2}{\eta_c}} \Big]{}_2F_1 \Big(1,\frac{\eta_c-2}{\eta_c};\frac{2\eta_c-2}{\eta_c};-s \rho_c T_d \Big)  }{ (2\pi \mathcal{U}_d)^{-1} (\rho_c T_d)^{\frac{2}{\eta_c}-1} (\eta_c-2) } \Bigg) \\
& \mathcal{L}_{\mathcal{I}_{cc}}(s)=\exp \Bigg(-\frac{ s \mathbb{E} \Big[ P_c^{\frac{2}{\eta_c}} \Big]{}_2F_1 \Big(1,\frac{\eta_c-2}{\eta_c};\frac{2\eta_c-2}{\eta_c};-s \rho_c \Big)  }{ (2\pi \lambda)^{-1} (\rho_c)^{\frac{2}{\eta_c}-1} (\eta_c-2) } \Bigg)\\
& \mathcal{L}_{\mathcal{I}_{ed}}(s)=\exp \Bigg(- \pi \mathcal{U}_e s^{\frac{2}{\eta_d}} \mathbb{E} \Big[ P_e^{\frac{2}{\eta_d}} \Big] \Gamma \Big(1+\frac{2}{\eta_d} \Big) \Gamma \Big(1-\frac{2}{\eta_d} \Big) \Bigg)\\
& \mathcal{L}_{\mathcal{I}_{dd}}(s)= \exp \Bigg(- \pi \mathcal{U}_d s^{\frac{2}{\eta_d}} \mathbb{E} \Big[ P_d^{\frac{2}{\eta_d}} \Big] \Gamma \Big(1+\frac{2}{\eta_d} \Big) \Gamma \Big(1-\frac{2}{\eta_d} \Big) \Bigg) \\
& \mathcal{L}_{\mathcal{I}_{cd}}(s)= \exp \Bigg(- \pi \lambda s^{\frac{2}{\eta_d}} \mathbb{E} \Big[ P_c^{\frac{2}{\eta_d}} \Big] \Gamma \Big(1+\frac{2}{\eta_d} \Big) \Gamma \Big(1-\frac{2}{\eta_d} \Big) \Bigg)
\end{align*}
and $\mathcal{L}_{\mathcal{I}_{ee}}(s)=\mathcal{L}_{\mathcal{I}_{ed}}(s),  \mathcal{L}_{\mathcal{I}_{de}}(s)=\mathcal{L}_{\mathcal{I}_{dd}}(s),  \mathcal{L}_{\mathcal{I}_{ce}}(s)=\mathcal{L}_{\mathcal{I}_{cd}}(s)$.
\normalsize
\textit{Proof:} See Appendix G. \qed

An important scenario of interest is the case of $\eta_c=4$, which does not only simplify the analysis but also represents a practical value for outdoor cellular communications in urban environments \cite{3B1,3B2,3B3,3B5,3B6,di_renzo}.

\textbf{\textit{Corollary 1}:} {For the \emph{network of interest},} at path-loss exponent $\eta_c=4$, the LTs of the interferences the cellular UEs experience from each communication mode reduce to:
\small
\begin{align*}
& \mathcal{L}_{\mathcal{I}_{ec}}(s) \stackrel{\eta_c=4}{=}\exp \Bigg(- \pi \mathcal{U}_e \sqrt{s} \mathbb{E} \Big[ \sqrt{P_e}\Big] \arctan(\sqrt{s \rho_c T_d})  \Bigg)\\
& \mathcal{L}_{\mathcal{I}_{dc}}(s) \stackrel{\eta_c=4}{=} \exp \Bigg(- \pi \mathcal{U}_d \sqrt{s} \mathbb{E} \Big[ \sqrt{P_d}\Big] \arctan(\sqrt{s \rho_c T_d})  \Bigg)\\
& \mathcal{L}_{\mathcal{I}_{cc}}(s) \stackrel{\eta_c=4}{=} \exp \Bigg(- \pi \lambda \sqrt{s} \mathbb{E} \Big[ \sqrt{P_c}\Big] \arctan(\sqrt{s \rho_c})  \Bigg). 
\end{align*}
\normalsize
\textit{Proof:} Simplify the expressions in Lemma 6 at $\eta_c=4$.\qed\\
Using the LTs of the interference PDFs, the outage probabilities of the D2D and cellular links are given in the following theorem:

\textbf{\textit{Theorem 1}:} {For the \emph{network of interest} }{with residual SI fraction $\zeta$,} the success probability for a UE operating in the cellular mode is given by \eqref{out_c} in general and by \eqref{out_c4} when $\eta_c=4$, the success probability for a UE operating in the f-D2D mode is given by \eqref{out_d}, and the success probability for a UE operating in the r-D2D mode is given by \eqref{out_e}.

\textit{Proof:} Using \eqref{BlahSINR_eqnGen} and the LTs of the interferences found in Lemma 6, we obtain the SINR outage probability expressions for each mode of operation. \qed

 \begin{figure*}[t]
\small
\begin{align}\label{out_c}
&\mathbb{P}(SINR_{c} \geq \theta)=
e^{   -\frac{\theta}{\rho_{c} } \sigma^2  - 2\pi (\frac{\theta}{\rho_{c} })^{\frac{2}{\eta_c}} \Bigg( \frac{{}_2F_1 \Big(1,\frac{\eta_c-2}{\eta_c};\frac{2\eta_c-2}{\eta_c};-\theta \Big)}{(\lambda  \mathbb{E} [ P_c^{\frac{2}{\eta_c}} ])^{-1}(\eta_c-2) \theta^{\frac{2}{\eta_c}-1}}    +    \frac{{}_2F_1 \Big(1,\frac{\eta_c-2}{\eta_c};\frac{2\eta_c-2}{\eta_c};-\theta T_d \Big)}{(\mathcal{U}_d \mathbb{E} [ P_d^{\frac{2}{\eta_c}} ])^{-1}(\eta_c-2) (\theta T_d)^{\frac{2}{\eta_c}-1}}    + \frac{{}_2F_1 \Big(1,\frac{\eta_c-2}{\eta_c};\frac{2\eta_c-2}{\eta_c};-\theta T_d \Big)}{(\mathcal{U}_e \mathbb{E} [ P_e^{\frac{2}{\eta_c}} ])^{-1}(\eta_c-2) (\theta T_d)^{\frac{2}{\eta_c}-1}}  \Bigg)   }  \\
&\stackrel{\eta_c=4}{=} \label{out_c4}
e^{   -\frac{\theta}{\rho_{c} } \sigma^2  - \pi \sqrt{\frac{\theta}{\rho_{c} }} \Big(\lambda  \mathbb{E} [ \sqrt{P_c} ] \arctan(\sqrt{\theta} )  +   \mathcal{U}_d \mathbb{E} [ \sqrt{P_d} ] \arctan(\sqrt{\theta T_d} ) + \mathcal{U}_e \mathbb{E} [ \sqrt{P_e}] \arctan(\sqrt{\theta T_d} )      \Big)   }\\
&\mathbb{P}(SINR_{d} \leq \theta)=  \label{out_d}
{  \mathbb{E}[e^{-\frac{\theta \zeta P_d }{\rho_d \mathbbm{1}_{FD}^{-1}}}]} e^{  - \frac{\theta}{\rho_{d} } \sigma^2  - \pi (\frac{\theta}{\rho_{d} } )^{\frac{2}{\eta_d}} \Big(\lambda  \mathbb{E} [ P_c^{\frac{2}{\eta_d}} ] {\Gamma(1+\frac{2}{\eta_d})\Gamma(1-\frac{2}{\eta_d})}     +   \mathcal{U}_d \mathbb{E} [ P_d^{\frac{2}{\eta_d}} ] {\Gamma(1+\frac{2}{\eta_d})\Gamma(1-\frac{2}{\eta_d})} + \mathcal{U}_e \mathbb{E} [ P_e^{\frac{2}{\eta_d}} ] {\Gamma(1+\frac{2}{\eta_d})\Gamma(1-\frac{2}{\eta_d})}      \Big)   } \\
&\mathbb{P}(SINR_{e} \leq \theta)=  \label{out_e}
{  \mathbb{E}[e^{-\frac{\theta \zeta P_e }{\rho_e \mathbbm{1}_{FD}^{-1} }}]} e^{  - \frac{\theta}{\rho_{e} } \sigma^2  - \pi (\frac{\theta}{\rho_{e} } )^{\frac{2}{\eta_d}} \Big(\lambda  \mathbb{E} [ P_c^{\frac{2}{\eta_d}} ] {\Gamma(1+\frac{2}{\eta_d})\Gamma(1-\frac{2}{\eta_d})}     +   \mathcal{U}_d \mathbb{E} [ P_d^{\frac{2}{\eta_d}} ] {\Gamma(1+\frac{2}{\eta_d})\Gamma(1-\frac{2}{\eta_d})} + \mathcal{U}_e \mathbb{E} [ P_e^{\frac{2}{\eta_d}} ] {\Gamma(1+\frac{2}{\eta_d})\Gamma(1-\frac{2}{\eta_d})}      \Big)   }
\end{align}
\normalsize
\end{figure*}
Let $\xi(t)=\frac{(e^t - 1)}{\rho_{\chi} }$; the link spectrum efficiency for a UE operating in mode $\chi$ is given by $\mathcal{R}_{\chi}$,
\footnotesize
\begin{align}
& \mathcal{R}_{\chi} = \mathbb{E}[\ln(1+SINR_{\chi})] = \int_0^{\infty} \mathbb{P} (\ln(1+SINR_{\chi}) > t) dt \notag \\
&= \int_0^{\infty}  e^{- \xi(t) \sigma^2}  { \mathcal{L}_{P_{\chi}} \Big( \xi(t) \zeta \mathbbm{1}_{FD}   \Big)} \underset{\kappa \in \{c,d,e\}}\prod \mathcal{L}_{\mathcal{I}_{\kappa \chi}} \Big( \xi(t)  \Big) dt.
\label{the_rates}
\end{align}
\normalsize
{In particular, the link spectrum efficiencies are,
\footnotesize
\begin{align*}
\mathcal{R}_{c} &= \int_0^{\infty}  e^{- \xi(t) \sigma^2}    \mathcal{L}_{\mathcal{I}_{c c}} \big( \xi(t)  \big) \mathcal{L}_{\mathcal{I}_{d c}} \big( \xi(t)  \big) \mathcal{L}_{\mathcal{I}_{e c}} \big( \xi(t)  \big) \; dt  \\
\mathcal{R}_{d} &= \int_0^{\infty}  e^{- \xi(t) \sigma^2}  { \mathcal{L}_{P_{d}} \big( \xi(t) \zeta \mathbbm{1}_{FD}   \big)} \mathcal{L}_{\mathcal{I}_{c d}} \big( \xi(t)  \big) \mathcal{L}_{\mathcal{I}_{d d}} \big( \xi(t)  \big) \mathcal{L}_{\mathcal{I}_{e d}} \big( \xi(t)  \big) \; dt  \\
& \stackrel{\zeta=0}{=} \int_0^{\infty}  e^{- \xi(t) \sigma^2}  \mathcal{L}_{\mathcal{I}_{c d}} \big( \xi(t)  \big) \mathcal{L}_{\mathcal{I}_{d d}} \big( \xi(t)  \big) \mathcal{L}_{\mathcal{I}_{e d}} \big( \xi(t)  \big) \; dt \\
  \mathcal{R}_{e} &= \int_0^{\infty}  e^{- \xi(t) \sigma^2}  { \mathcal{L}_{P_{e}} \big( \xi(t) \zeta \mathbbm{1}_{FD}   \big)} \mathcal{L}_{\mathcal{I}_{c e}} \big( \xi(t)  \big) \mathcal{L}_{\mathcal{I}_{d e}} \big( \xi(t)  \big) \mathcal{L}_{\mathcal{I}_{e e}} \big( \xi(t)  \big) \; dt  \\
&\stackrel{\zeta=0}{=} \int_0^{\infty}  e^{- \xi(t) \sigma^2}  \mathcal{L}_{\mathcal{I}_{c e}} \big( \xi(t)  \big) \mathcal{L}_{\mathcal{I}_{d e}} \big( \xi(t)  \big) \mathcal{L}_{\mathcal{I}_{e e}} \big( \xi(t)  \big) \; dt. 
\end{align*} }
\normalsize
The SINR CDFs and the spectral efficiencies are the core contributions of this paper which allow us to analyze the cellular network with FD-enabled D2D. Theorem 1 is validated against system level simulations in the next section.

\section{Results and Analysis}\label{Results}

In this section, we validate the developed mathematical model and benchmark the FD-D2D operation against the D2D enabled cellular network with HD UEs, denoted as the HD-network, and the traditional cellular network where D2D is disabled. The cellular network overlaid with FD-D2D being considered in this work will be referred to as the FD-network. Let $\mathcal{A}_\chi$ be the probability of the joint event that a randomly selected user is operating in mode $\chi$ and is not in truncation outage, $\text{PUR}(\chi)$ and $\text{PCR}(\chi)$ be the average per-user rate and average per-cell rate, respectively, of mode $\chi$, $\Lambda(\chi)$ be the intensity of users operating in mode $\chi$, and $\text{Tx}(\chi)$ be the average transmit power of users operating in mode $\chi$, where the value of each of these parameters in each network scenario is given in Table~\ref{Modes_Table}. Note that $\text{PUR}(c)$ in Table~\ref{Modes_Table} has a multiplication factor of $\frac{1}{2} \beta$ in which the factor $\frac{1}{2}$ reflects the two-hop nature (i.e. uplink then downlink) of the cellular links and the factor $\beta=\frac{\text{BS intensity}}{\text{intensity of UEs in cellular mode}} = \frac{\lambda}{(1-\mathcal{O}_p)\lambda_c}$ reflects the share each user get from the uplink spectrum when equal sharing among the users is assumed. On the other hand, $\text{PCR}(c)$ does not incorporate the two hops or the spectrum sharing factors because we look at the total uplink rate from the BS side. Assuming a round robin scheduling for the cellular UEs, {$\text{Tx}(c)$ is also multiplied with the factor $\beta$ to reflect the activity of the UEs.}

As shown in Table~\ref{Modes_Table}, cellular and f-D2D links share the spectrum in the HD-network, while the spectrum is explicitly used by cellular links in the traditional cellular case. Therefore, the rates in Table~\ref{Modes_Table} are explicitly defined for each network scenario to reflect their different interference environments, where $\mathcal{R}^{(\text{Conv})}_{\chi} > \mathcal{R}^{{(\text{HD})}}_{\chi} > \mathcal{R}^{{(\text{FD})}}_{\chi}$. Note that $ \mathcal{R}^{{(\text{FD})}}_{\chi}$ is given in \eqref{the_rates}. The rates $\mathcal{R}^{{(\text{HD})}}_{\chi}$ and $\mathcal{R}^{(\text{Conv})}_{\chi}$ are evaluated via \eqref{the_rates} by eliminating the LT of the SI along with $\mathcal{L}_{\mathcal{I}_{e \chi}} (\cdot)$ and $\mathcal{L}_{\mathcal{I}_{d \chi}}(\cdot) \mathcal{L}_{\mathcal{I}_{e \chi}}(\cdot)$, respectively.

From the user side, we define two performance metrics to assess the per-user gain in the FD-network when compared to the HD-network and traditional network. The first metric is the per-user rate, defined as  $\mathcal{T}_{avg}= \sum_{\chi } \mathcal{A}_{\chi} \text{PUR}({\chi})$. The second metric is the average transmit power, defined as $P_{avg}= \sum_{\chi } \mathcal{A}_{\chi} \text{Tx}({\chi})$. The performance gain from the network side is evaluated by the network throughput, which is defined as $\mathcal{T}_n= \lambda  \sum_{\chi}    \text{PCR}({\chi})$.

\footnotesize
\begin{table}[t]
\centering
\small
\caption{{\color{black} Mode of Operation Parameters} }
\resizebox{0.49 \textwidth}{!}{\begin{tabular}{|c|c|c|c|c|c|c|}
\hline
\multirow{-2}{*}{}
& & &&& &  \\
\multirow{-2}{*}{\large \textbf{Network}}&  \multirow{-2}{*}{\large\textbf{Mode $(\chi)$}}&  \multirow{-2}{*}{\large\textbf{$\mathcal{A}_\chi$}}&  \multirow{-2}{*}{\large \textbf{\text{PUR}$(\chi)$ }} &  \multirow{-2}{*}{\large \textbf{\text{PCR}$(\chi)$ }}  & \multirow{-2}{*}{ \textbf{$\Lambda(\chi)$} }&  \multirow{-2}{*}{\textbf{Tx$(\chi)$} } \\ \hline \hline

 \multirow{6}{*}{\large \textbf{FD-network}} & \multirow{2}{*}{\large Cellular $(c)$} &  \multirow{2}{*}{\large $ \frac{\lambda_c}{2\lambda_d+\lambda_c}(1-\mathcal{O}_p)  $} & \multirow{2}{*}{\large $\frac{1}{2} \beta {\mathcal{R}^{(\text{FD})}_c}$}  &   \multirow{2}{*}{\large $\mathcal{R}^{(\text{FD})}_c $} &\multirow{2}{*}{\large $ \lambda $} & \multirow{2}{*}{\large $\beta \mathbb{E}[P_c] $} \\
 & &  &   &  & & \\ \cline{2-7}
 & \multirow{2}{*}{\large f-D2D $(d)$} & \multirow{2}{*}{\large $ \frac{\lambda_d}{2\lambda_d+\lambda_c}  \mathcal{P}_d  $} &   \multirow{2}{*}{\large $ {\mathcal{R}^{(\text{FD})}_d} $}  &   \multirow{2}{*}{\large $ \frac{\mathcal{U}_d}{\lambda}\mathcal{R}^{(\text{FD})}_d $}  & \multirow{2}{*}{\large $\mathcal{U}_d $} & \multirow{2}{*}{\large $\mathbb{E}[P_d] $} \\   
& & &&& & \\  \cline{2-7}
 & \multirow{2}{*}{\large r-D2D $(e)$} &  \multirow{2}{*}{\large $ \frac{\lambda_d}{2\lambda_d+\lambda_c}  \mathcal{P}_e  $}  & \multirow{2}{*}{\large $  {\mathcal{R}^{(\text{FD})}_e} $}    &   \multirow{2}{*}{\large $\frac{\mathcal{U}_e}{\lambda} \mathcal{R}^{(\text{FD})}_e $} & \multirow{2}{*}{\large $\mathcal{U}_e $} & \multirow{2}{*}{\large $\mathbb{E}[P_e] $}    \\
 & & &&& &  \\   \hline
  \hline

  \multirow{4}{*}{\large \textbf{HD-network}} & \multirow{2}{*}{\large Cellular $(c)$} &  \multirow{2}{*}{\large $ \frac{\lambda_c}{\lambda_d+\lambda_c} (1-\mathcal{O}_p)  $} & \multirow{2}{*}{\large $\frac{1}{2} \beta {\mathcal{R}^{(\text{HD})}_c}$}  &   \multirow{2}{*}{\large $\mathcal{R}^{(\text{HD})}_c $} &\multirow{2}{*}{\large $ \lambda $} & \multirow{2}{*}{\large $\beta \mathbb{E}[P_c] $} \\
 & &  &   &  & & \\ \cline{2-7}
  & \multirow{2}{*}{\large f-D2D $(d)$} & \multirow{2}{*}{\large $ \frac{\lambda_d}{\lambda_d+\lambda_c}  \mathcal{P}_d $} &   \multirow{2}{*}{\large $ {\mathcal{R}^{(\text{HD})}_d} $}  &   \multirow{2}{*}{\large $ \frac{\mathcal{U}_d}{\lambda}\mathcal{R}^{(\text{HD})}_d $}  & \multirow{2}{*}{\large $\mathcal{U}_d $} & \multirow{2}{*}{\large $\mathbb{E}[P_d] $} \\   
& & &&& & \\  \hline \hline
 
   \multirow{2}{*}{\large \textbf{Traditional network}}& \multirow{2}{*}{\large Cellular $(c)$} &  \multirow{2}{*}{\large $ (1-\mathcal{O}_p) $} & \multirow{2}{*}{\large $\frac{1}{2} \beta {\mathcal{R}^{(\text{Conv.})}_c}$}  &   \multirow{2}{*}{\large $\mathcal{R}^{(\text{Conv.})}_c $} &\multirow{2}{*}{\large $ \lambda $} & \multirow{2}{*}{\large $\beta \mathbb{E}[P_c] $} \\
 & &  &   &  & & \\ \hline

\end{tabular}}
\label{Modes_Table}
\end{table}
\normalsize

\subsection{Parameter Selection}
\begin{table}[t]
\centering
\small
\caption{{\color{black} Parameter Values} }
\resizebox{0.28 \textwidth}{!}{\begin{tabular}{|c|c|c|c|}
\hline
\multirow{-2}{*}{}
& & &   \\
\multirow{-2}{*}{\textbf{\large Parameter}}&  \multirow{-2}{*}{\textbf{\large Value}}&  \multirow{-2}{*}{\textbf{\large Parameter}}&  \multirow{-2}{*}{\textbf{\large Value}} \\
 & & &  \\ \hline
   \multirow{2}{*}{\textbf{\large $\lambda$}}& \multirow{2}{*}{\large 10 BS/km$^2$}&    \multirow{2}{*}{\textbf{\large $\rho_d$ }}& \multirow{2}{*}{\large $\frac{\rho_c}{r_1}$  }  \\
 & & &  \\ \hline
    \multirow{2}{*}{\textbf{\large $\lambda_c$}}& \multirow{2}{*}{\large 100 UE/km$^2$}& \multirow{2}{*}{\textbf{\large $\rho_e$}}& \multirow{2}{*}{\large $\frac{\rho_d}{r_2}=\frac{\rho_c}{r_1 r_2} $ }  \\
& & &  \\ \hline
     \multirow{2}{*}{\textbf{\large $\lambda_d$}}& \multirow{2}{*}{\large 100 UE/km$^2$}& \multirow{2}{*}{\textbf{\large $\eta_c$}}& \multirow{2}{*}{\large 4}   \\
 & & &  \\ \hline
     \multirow{2}{*}{\textbf{\large $P_u$}}& \multirow{2}{*}{\large 200 mW}& \multirow{2}{*}{\textbf{\large $\eta_d$ }}& \multirow{2}{*}{\large 4 } \\
 & & &  \\ \hline
    \multirow{2}{*}{\textbf{\large $\rho_{min}$}}& \multirow{2}{*}{\large -90 dBm}& \multirow{2}{*}{\textbf{\large $\omega$}}& \multirow{2}{*}{\large 1}  \\
& & &  \\ \hline
    \multirow{2}{*}{\textbf{\large $\rho_c$}}& \multirow{2}{*}{\large -80 dBm}& \multirow{2}{*}{\textbf{\large $\sigma^2$}}& \multirow{2}{*}{\large -90 dBm}  \\
 & & &  \\ \hline
\end{tabular}}
\label{Parameter_Table}
\end{table}
\normalsize
{In Sections \ref{Results}-\emph{B}$-$\ref{Results}-\emph{E}}, we focus on the FD-network performance assuming perfect SIC (i.e., $\zeta=0$) to study the explicit contribution of the FD communication to the aggregate interference and the subsequent effect on outage and rate. Once FD-D2D gains over the HD and traditional networks are highlighted, the effect of imperfect SIC is studied (i.e., $\zeta>0$) {in Section \ref{Results}-\emph{F}. Section \ref{Results}-\emph{G} focuses on the effect of the link distance distribution and so assumes perfect SIC.} {For the simulation scenario, unless stated otherwise, the parameter values in Table \ref{Parameter_Table} are used. Proposition 1 is used for the distribution of $f_{r_e}(x)$ in the analysis.} 


{For a fixed $\rho_c$, we control our network using three main parameters: $r_1$, $r_2$, and $T_d$. In particular, $r_1$ controls the power required at both the f-D2D and r-D2D UEs; decreasing $r_1$  implies higher $\rho_d$ and $\rho_e$ at the receiver UEs. Using $r_2$ we control the power required at the r-D2D link's receiver only; decreasing $r_2$ implies higher $\rho_e$ at the f-D2D UEs. The amount of IP provided to BSs is controlled by $T_d$ ; increasing $T_d$ loosens the IP conditions by increasing the maximum allowed interference  (i.e., $T_d \rho_c$), which allows more f-D2D and r-D2D UEs to satisfy the mode selection inequalities and transmit.} {Table \ref{blah_Table} summarizes the effects of varying $r_1$, $r_2$, and $T_d$ on the network, where ($\uparrow$), ($\downarrow$), and ($-$) denote increase, decrease, and no change, respectively.} 

\begin{table}[t]
\centering
\caption{{\color{black} Effect of increasing $r_1$, $r_2$, and $T_d$.} }
\resizebox{0.48 \textwidth}{!}{\begin{tabular}{|c|c|c|c|c|c|}
\hline
 &  \multicolumn{2}{ c }{ f-D2D}  & \multicolumn{2}{ |c| }{r-D2D } & Cellular \\ 
 \cline{2-6}
 \multirow{-2}{*}{}
& & && &   \\
\multirow{-3}{*}{\textbf{Parameter }}&  \multirow{-2}{*}{$\mathcal{U}_d$}&  \multirow{-2}{*}{$\rho_d$} &  \multirow{-2}{*}{ $\mathcal{U}_e$} &  \multirow{-2}{*}{ $\rho_e$}  & \multirow{-2}{*}{IP} \\ 
\hline
 \multirow{2}{*}{\textbf{Increasing $r_1$}} & \multirow{2}{*}{$\uparrow$} &  \multirow{2}{*}{$\downarrow$}  &  \multirow{2}{*}{$\uparrow$}  &   \multirow{2}{*}{$\downarrow$}  & \multirow{2}{*}{$-$}\\
 
& &  &   & & \\ \hline

  \multirow{2}{*}{\textbf{Increasing $r_2$ }} & \multirow{2}{*}{$-$}  &\multirow{2}{*}{$-$}  &   \multirow{2}{*}{$\uparrow$}  &   \multirow{2}{*}{$\downarrow$} &  \multirow{2}{*}{$-$} \\
  
   & &  &   &  &   \\ \cline{1-4}  \cline{5-6}
 
   \multirow{2}{*}{\textbf{Increasing $T_d$ }}& \multirow{2}{*}{$\uparrow$} &  \multirow{2}{*}{$-$}    &  \multirow{2}{*}{$\uparrow$}  &   \multirow{2}{*}{$-$}   & \multirow{2}{*}{$\downarrow$}  \\
   & &  &   &   & \\ \hline

\end{tabular}}
\label{blah_Table}
\end{table}

\subsection{Model Validation}
We first validate Approximations 1, 2, 3, and 4, as well as Proposition 1 for an FD-network by showing that the analysis is a good estimate of the simulations. Note that the simulation does not enforce any of these approximations. \textcolor{black}{In each simulation run, a PPP cellular network with BS intensity $\lambda$ is simulated in a 1000 km$^2$ area. We then generate cellular and f-D2D UEs with intensities $\lambda_c$ and $\lambda_d$, respectively. For each f-D2D UE, an r-D2D UE is generated within a radius of $\bar{R}$ according to the PDF in \eqref{power_distance}. The cellular UEs are scheduled to transmit if they are not in truncation and if there is no other UE scheduled in the same Voronoi cell. The f-D2D and r-D2D UEs are scheduled to transmit if they satisfy the maximum transmit power constraint and IP. All UEs employ channel inversion power control.} 

Fig. \ref{changing_r2} is a plot of SINR outage against $\theta$ with $r_1$ fixed to 1 (i.e. $\rho_d = \rho_c$). The figure shows that increasing $r_2$ (i.e. decreasing $\rho_e$) worsens SINR for all three communication modes. This occurs due to the increase in the number of transmitting r-D2D UEs which increases network interference. The r-D2D mode, however, is impacted significantly more than the cellular and f-D2D modes as increasing $r_2$ not only increases interference but also worsens the received signal power ($\rho_e$) of the r-D2D mode.

Fig. \ref{changing_r1andTd} shows the SINR outage of all three communication modes increases when $r_1$ is increased (for fixed $r_2$ and $T_d$) and when $T_d$ is increased (for fixed $r_1$ and $r_2$). Increasing $r_1$ and $T_d$ each increases outage due to the increased interference that results. However, $r_1$ impacts the f-D2D and r-D2D modes more significantly than the cellular mode as it additionally affects their received signal powers. Since varying $T_d$ only affects the network interference, the impact on the SINR of all three communication modes is similar. This difference in sensitivity to the parameters $r_1$ and $r_2$ gives the network operator the ability to alter the respective parameter if the performance of one of these modes needs to be altered without affecting the other modes too much, thereby giving flexibility to control the performance of a mode without affecting the other modes significantly.

\begin{figure}
\begin{minipage}[t]{0.9\linewidth}
\centering\includegraphics[scale=0.5]{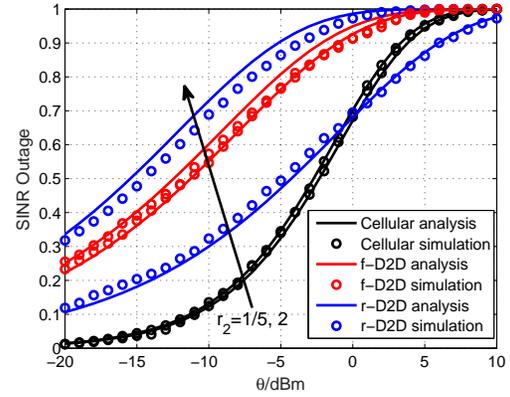}
\caption{SINR Outage vs. $\theta$ for $\rho_c=-80$dBm, $r_1=1$ and $T_d=0.2$ and different $r_2$.}\label{changing_r2}
\end{minipage}
\end{figure}

\begin{figure}
\begin{minipage}[t]{0.9\linewidth}
\centering\includegraphics[scale=0.35]{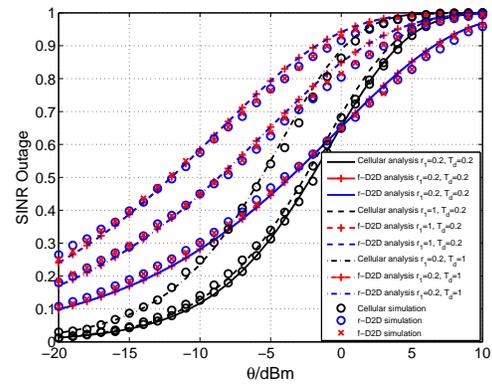}
\caption{SINR Outage vs. $\theta$ for $\rho_c=-80$dBm, $r_2=1$, and different $r_1$ and $T_d$.}\label{changing_r1andTd}
\end{minipage}
\end{figure}

\subsection{Effects of the Interference Protection Condition}
Increasing $T_d$ decreases the amount of IP for the cellular mode and thereby allows a larger number of D2D UEs to transmit. This increases spatial frequency reuse but also increases network interference. Hence, $T_d$ imposes a tradeoff between the number of simultaneously active links and the transmission rate per link.

{We first show the network throughput vs $T_d$ for the FD-network and HD-networks normalized w.r.t. the traditional cellular network in Fig. \ref{vs_Td_Tn}. Note that $\mathcal{T}_n$ of the traditional cellular network does not change with $T_d$ as D2D communication is prohibited. The figure shows the existence of an optimal $T_d$ that maximizes the network throughput. An optimal $T_d$ exists because increasing $T_d$, at first, has a larger positive impact on the overall performance by increasing the number of transmitting D2D UEs, thereby increasing spatial frequency reuse. Beyond the optimal $T_d$, the negative impact of the D2D interference dominates the network performance and increasing $T_d$ deteriorates the network throughput. An important observation from the figure is that the FD-network offers non-trivial throughput gains when compared to the HD-network and traditional cellular network, {64\% and 245\%}, respectively. These high gains are observed because we look at the performance from the network perspective in which the FD-network allows an additional $\mathcal{U}_e$ and additional ($\mathcal{U}_d + \mathcal{U}_e$) links per unit area to efficiently reuse the spectrum when compared to the HD-network and traditional network, respectively. It is also worth mentioning that the HD-network allows an additional $\mathcal{U}_d$ links per unit area to reuse the spectrum compared to the traditional network, which gives {110\%} increase in the throughput.}

{The D2D bias factor $T_d$ also affects the average transmit power, $P_{avg}$, of the D2D devices as shown in Fig. \ref{vs_Td_avg_power}. The figure shows that D2D communication generally reduces the average transmission power when compared to the traditional cellular network for low values of $T_d$. This occurs because low $T_d$ only allows D2D UEs that have lower path-loss attenuation to transmit which reduces the transmission power due to the employed channel inversion power control. {Also, for lower $T_d$, the FD-network offers a lower power consumption than the HD-network because it allows a larger number of devices to exploit good channel conditions and communicate in the D2D mode. However, for larger $T_d$, UEs with higher transmit-powers are allowed to transmit; since, the FD-network allows a larger number of these than the HD-network, its average power consumption exceeds the HD-network's.} Interestingly, the $T_d$ that optimizes spectral efficiency (cf. Fig. \ref{vs_Td_Tn}) falls in the region that offers high transmit power reduction w.r.t. the traditional network, implying that using the correct value of $T_d$ enables the network to simultaneously consume less power per transmitting UE and gain maximum throughput.}

\begin{figure}
\begin{minipage}[t]{0.9\linewidth}
\centering\includegraphics[scale=0.5]{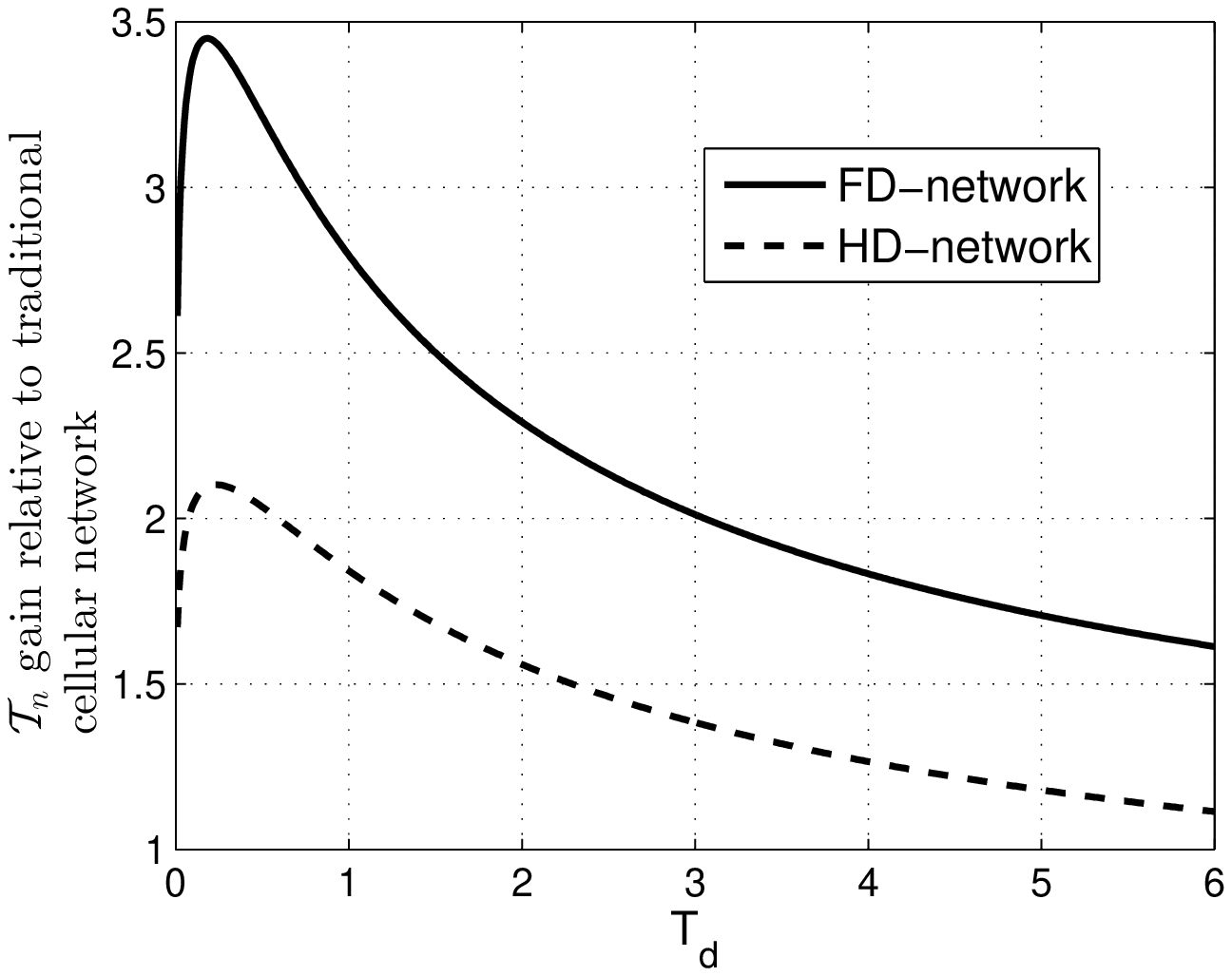}
\caption{$\mathcal{T}_n$ gain w.r.t the traditional cellular network vs. $T_d$ with $r_1=0.2$ and $r_2=0.2$.}\label{vs_Td_Tn}
\end{minipage}\\
\begin{minipage}[t]{0.9\linewidth}
\centering\includegraphics[scale=0.5]{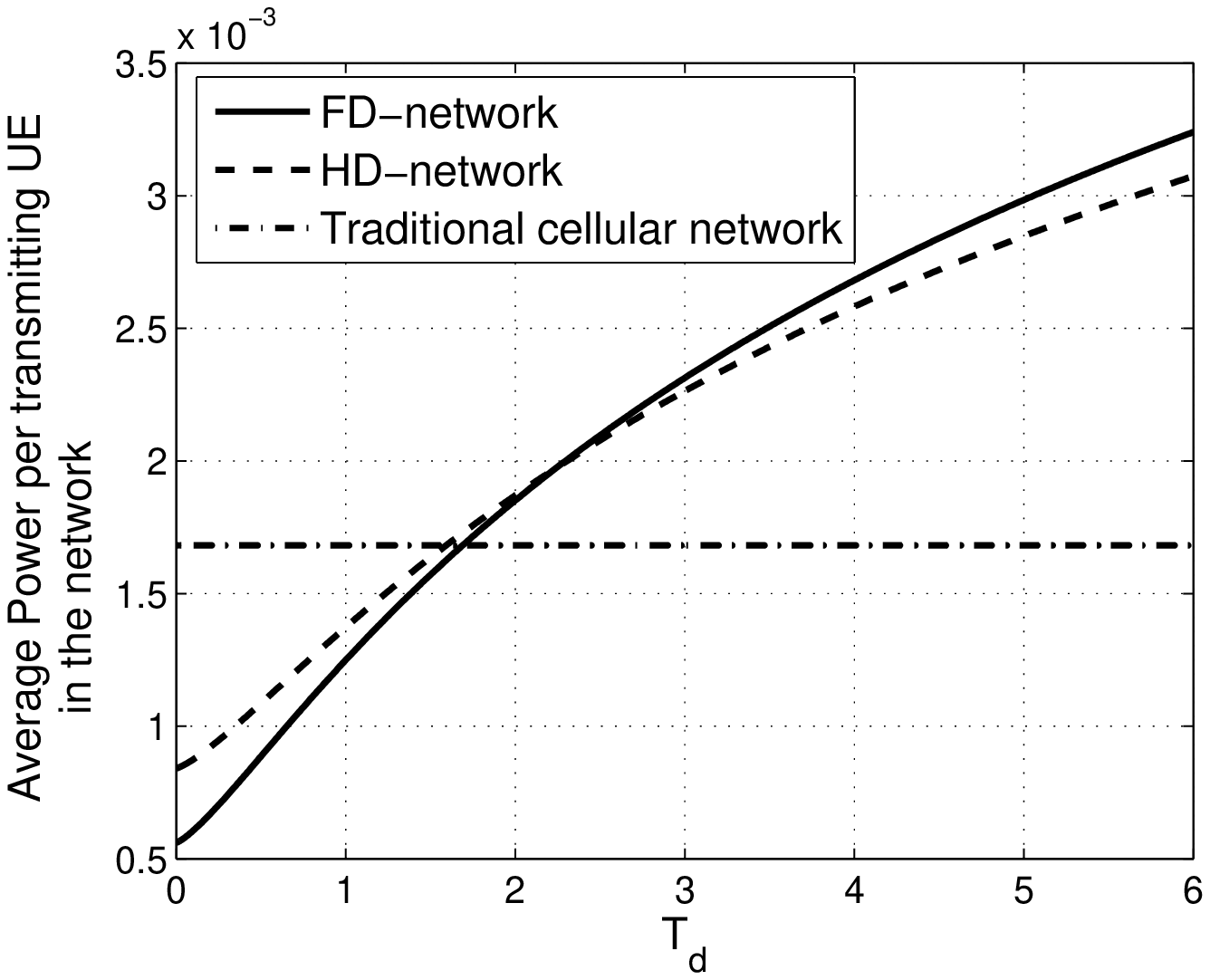}
\caption{$P_{avg}$ vs. $T_d$ with $r_1=0.2$ and $r_2=0.2$.}\label{vs_Td_avg_power}
\end{minipage}
\end{figure}

Fig. \ref{vs_Td_Outs_Sep} shows the negative impact of the increased D2D communication on the explicit SINR outage of each mode of operation for all values of $T_d$. In particular, for each mode of operation (cellular, f-D2D, and r-D2D), inducing more D2D communication deteriorates the SINR outage. Hence the SINR outage of common modes of operation is highest for the FD-network, followed by the HD-network, and finally the traditional cellular network. Also, the figure manifests the crucial role of IP on the cellular network outage probability and the drastic rate of outage increase with increasing $T_d$.

Fig. \ref{h2} is a plot of the average SINR network-outage against $T_d$, which is defined as {$\mathcal{O}_{net}= \sum_{\chi } \frac{\Lambda(\chi)}{{\sum_{\kappa }\Lambda(\kappa)}} \mathcal{O}_{\chi}$}. Note that we differentiate between the outage probabilities in each network scenario, namely, $\mathcal{O}^{(\text{FD})}_{\chi}$, $\mathcal{O}^{(\text{HD})}_{\chi}$, and $\mathcal{O}^{(\text{Conv.})}_{\chi}$, according to the interference environment. Hence, $ \mathcal{O}^{{(\text{FD})}}_{\chi}$ is given in \eqref{BlahSINR_eqnGen} and the outages $\mathcal{O}^{{(\text{HD})}}_{\chi}$ and $\mathcal{O}^{(\text{Conv})}_{\chi}$ are evaluated via \eqref{BlahSINR_eqnGen} by eliminating the LT of the SI along with $\mathcal{L}_{\mathcal{I}_{e \chi}} (\cdot)$ and $\mathcal{L}_{\mathcal{I}_{d \chi}} (\cdot) \mathcal{L}_{\mathcal{I}_{e \chi}}(\cdot)$, respectively. As shown in the figure, the FD-network has the highest network-outage, followed by the HD-network, and finally the traditional cellular network. The figure manifests the importance of $T_d$ and shows that the FD-network requires a much more stringent IP-condition to maintain the same outage performance as the HD-network.

\begin{figure} 
\begin{minipage}[t]{0.9\linewidth}
\centering\includegraphics[scale=0.5]{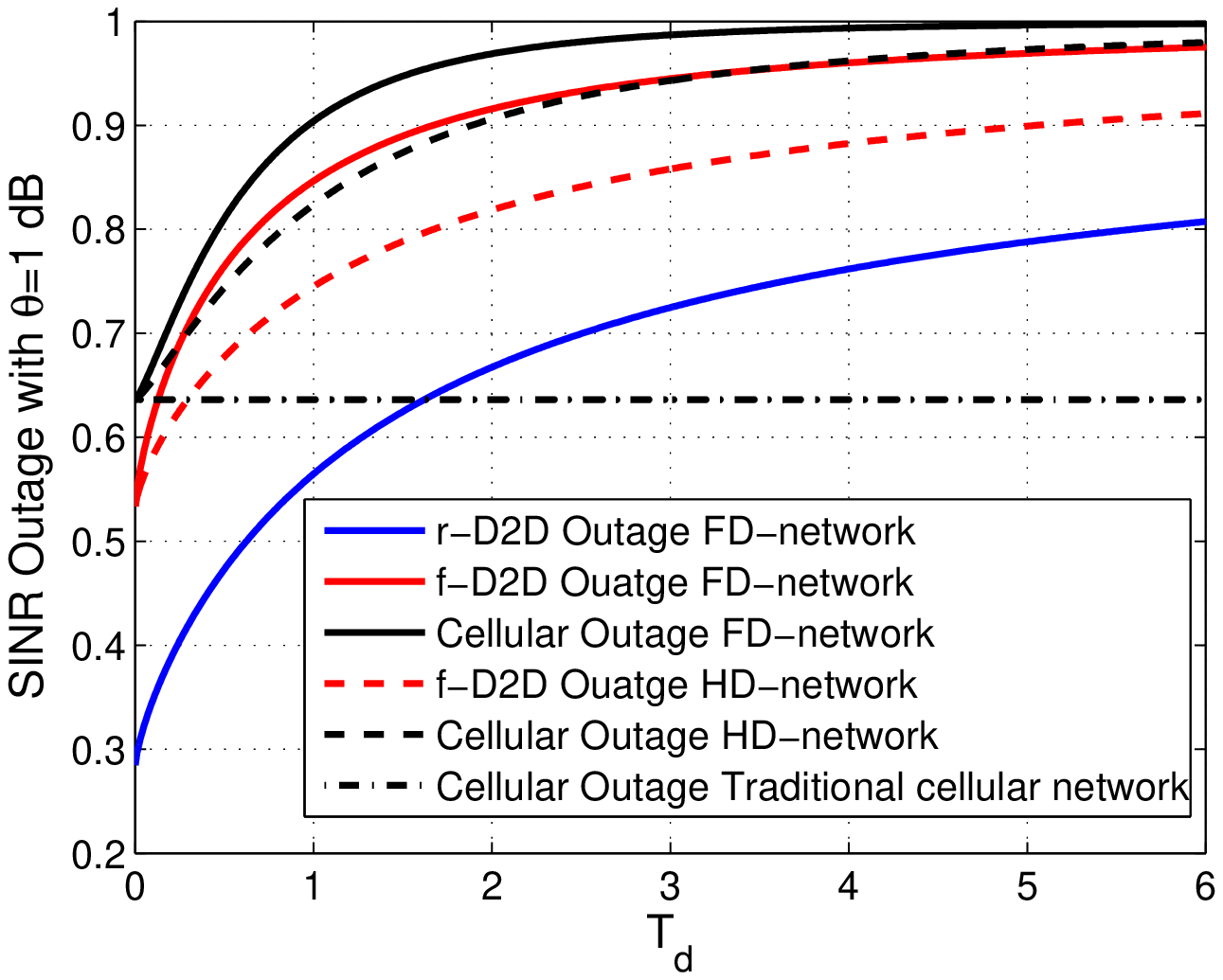}
\caption{SINR Outage vs. $T_d$ for $\theta=1$ with $r_1=0.2$ and $r_2=0.2$.}\label{vs_Td_Outs_Sep}
\end{minipage}\;\;\;
\begin{minipage}[htb]{0.9\linewidth}
\centering\includegraphics[scale=0.5]{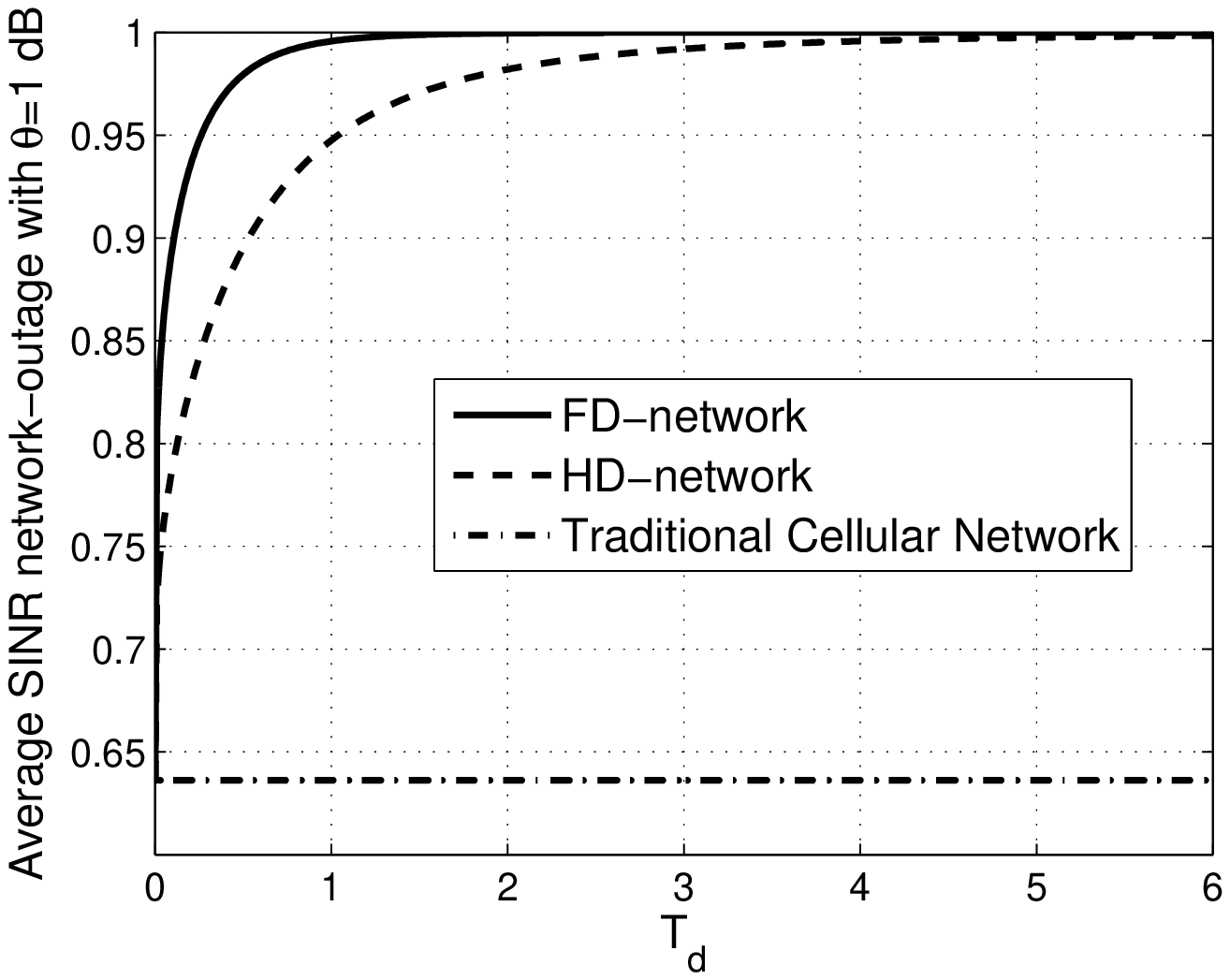}
\caption{$\mathcal{O}_{net}$ vs. $T_d$ for $\theta=1$ with $r_1=1$ and $r_2=2$.}\label{h2}
\end{minipage}
\end{figure}

Figs. \ref{vs_Td_Tn} and \ref{h2} clearly show the tradeoff between spectral efficiency and outage probability. It could be concluded that despite the increased outage probability, the overall network capacity increases due to the improved spatial frequency reuse. {It is worth mentioning that the high numerical values for outage probabilities in Figs. \ref{vs_Td_Outs_Sep} and \ref{h2} are common in stochastic geometry based analysis due to the simplified system model and the employed simplistic interference management scheme to maintain tractability. Nevertheless, despite the increased outage, FD-D2D communication provides potential gains to the per-user as well as the aggregate network throughputs in cellular networks. In practice, the ignored effect of shadowing and propagation along with employing more sophisticated interference management schemes are expected to reduce SINR outage and increase the harvested FD-D2D gains.}

%

%

\subsection{Effects of the Distance Cut-off for the FD Mode}
Increasing $r_2$ (i.e. decreasing $\rho_e$) decreases the power required by r-D2D UEs to invert their channel and thereby increases the number of transmitting r-D2D UEs. This increases spatial frequency reuse but also increases network interference. Additionally, the received intended signal power of the r-D2D links decreases.

Fig. \ref{vs_r2_Ravg} shows the existence of an optimal $r_2$ that maximizes $\mathcal{T}_{avg}$. An optimal exists because increasing $r_2$, is beneficial at first as it increases the number of transmitting r-D2D UEs that make a useful contribution to the network. Increasing $r_2$ beyond this causes deterioration to the overall performance due to the increased interference, as well as due to the decreased power of the intended signal of the r-D2D UEs (i.e., $\rho_e$).  
 \textcolor{black}{Fig. \ref{vs_r2_Ravg} shows that at the optimal $r_2$, the FD-network outperforms the HD-network by {18\%}. Both the HD-network and the FD-network outperform the traditional cellular network significantly.} {The high performance gain offered by the D2D communication (both FD-D2D and HD-D2D) w.r.t. the traditional cellular network can be attributed to the explicit utilization by each D2D link for the available uplink channel when compared to the share $\beta$ that UEs get when scheduled in the cellular mode. Furthermore, D2D communication establishes a direct (i.e., one-hop) link between two UEs compared to the two-hop (i.e., uplink then downlink) communication via the BS. }

\begin{figure}
\begin{minipage}[htb]{0.9\linewidth}
\centering\includegraphics[scale=0.5]{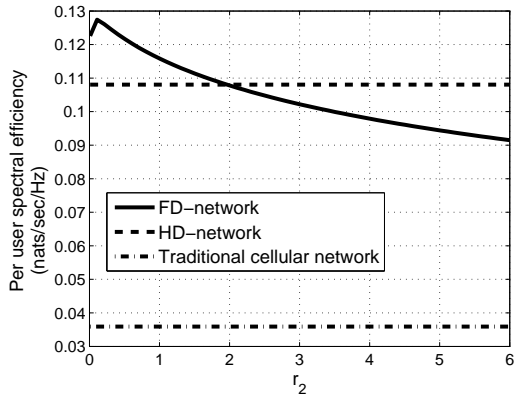}
\caption{$\mathcal{T}_{avg}$ vs. $r_2$ with $r_1=0.01$ and $T_d=1$.}\label{vs_r2_Ravg}
\end{minipage}\;\;\;
\end{figure}


\subsection{Analyzing Truncation and SINR Outage}


\begin{figure}
\begin{minipage}[htb]{0.9\linewidth}
\centering\includegraphics[scale=0.5]{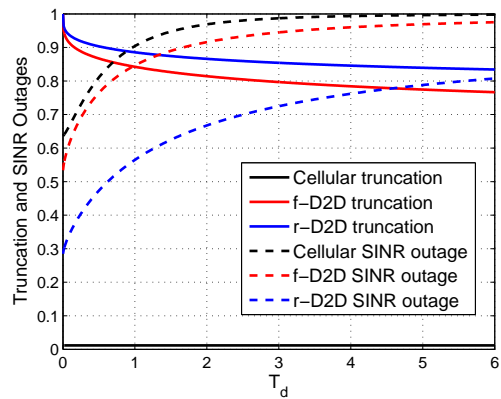}
\caption{Truncation and SINR Outages vs. $T_d$ with $r_1=0.2$ and $r_2=0.2$.}\label{vs_Td_truncAndSinrOuts}
\end{minipage}
\end{figure}

Fig. \ref{vs_Td_truncAndSinrOuts} is a plot of the truncation and SINR outages of the individual transmission modes with increasing $T_d$ for an FD-network. Note, the r-D2D links and f-D2D links have higher truncation outage than the cellular links due to the small values of $r_1$ and $r_2$ (and therefore high values of $\rho_d$ and $\rho_e$) being used. Our goal is to observe the effect of increasing $T_d$ on the truncation and SINR outages. Since increasing $T_d$ decreases IP, the truncation outages of the f-D2D and r-D2D transmission modes decrease with $T_d$ until they settle to a constant. This occurs when the inability to invert the channel to the receiver becomes the bottleneck of truncation outage and not the inability to comply with the IP-condition. At the same time we see that increasing $T_d$, which allows more f-D2D and r-D2D links, increases SINR outage. Increasing $T_d$ \textcolor{black}{allows more D2D transmissions that cause more interference to the BSs}; this occurs either when the links have high power and/or when the transmitting UE is closer to the BS. 

\subsection{{Effects of Imperfect SIC}}
{In this set of results, we investigate the effect of imperfect SIC on the FD-D2D network performance. First, we look at SINR outage probability for different values of the residual SI fraction $\zeta$ in Fig. \ref{out_all_zeta}. As expected, increasing $\zeta$ deteriorates the outage probability for f-D2D and r-D2D UEs due to the increased residual SI. {It ought to be highlighted that only a fraction $\mathcal{P}_{FD}$ of the f-D2D and r-D2D links operate in FD and experience residual SI. Additionally, in Fig. \ref{out_all_zeta}, we note that the r-D2D mode is impacted more severely by the residual SI than the f-D2D mode. This occurs because of the $r_2$ being used, which increases $\rho_e$ and therefore r-D2D transmission powers, which in turn leads to more SI for the r-D2D UEs.} 


We also look at the effect of imperfect SIC on the total network throughput in Figs. \ref{vsTd_Tn_zeta} and \ref{vsTd_Tn_zeta_good} for different values of $r_1$ and $r_2$. In both cases, the figures show that increasing $\zeta$ deteriorates the network throughput due to the imposed SI on the FD links. {The HD-network is included in the figures to benchmark the FD-D2D with imperfect SIC. Figs. \ref{vsTd_Tn_zeta} and \ref{vsTd_Tn_zeta_good} show a $\zeta$ dependent threshold where the HD-network outperforms the FD-network when $\zeta$ is high. In particular, Fig. \ref{vsTd_Tn_zeta_good} shows that if the network parameters are properly tuned, higher $\zeta$ values become more tolerable allowing the FD-network to outperform the HD-network. This highlights the importance of properly tuning the network parameters allowing, theoretically, the FD-network to always outperform the HD-network.}
\begin{figure}
\begin{minipage}[htb]{0.9\linewidth}
\centering\includegraphics[scale=0.5]{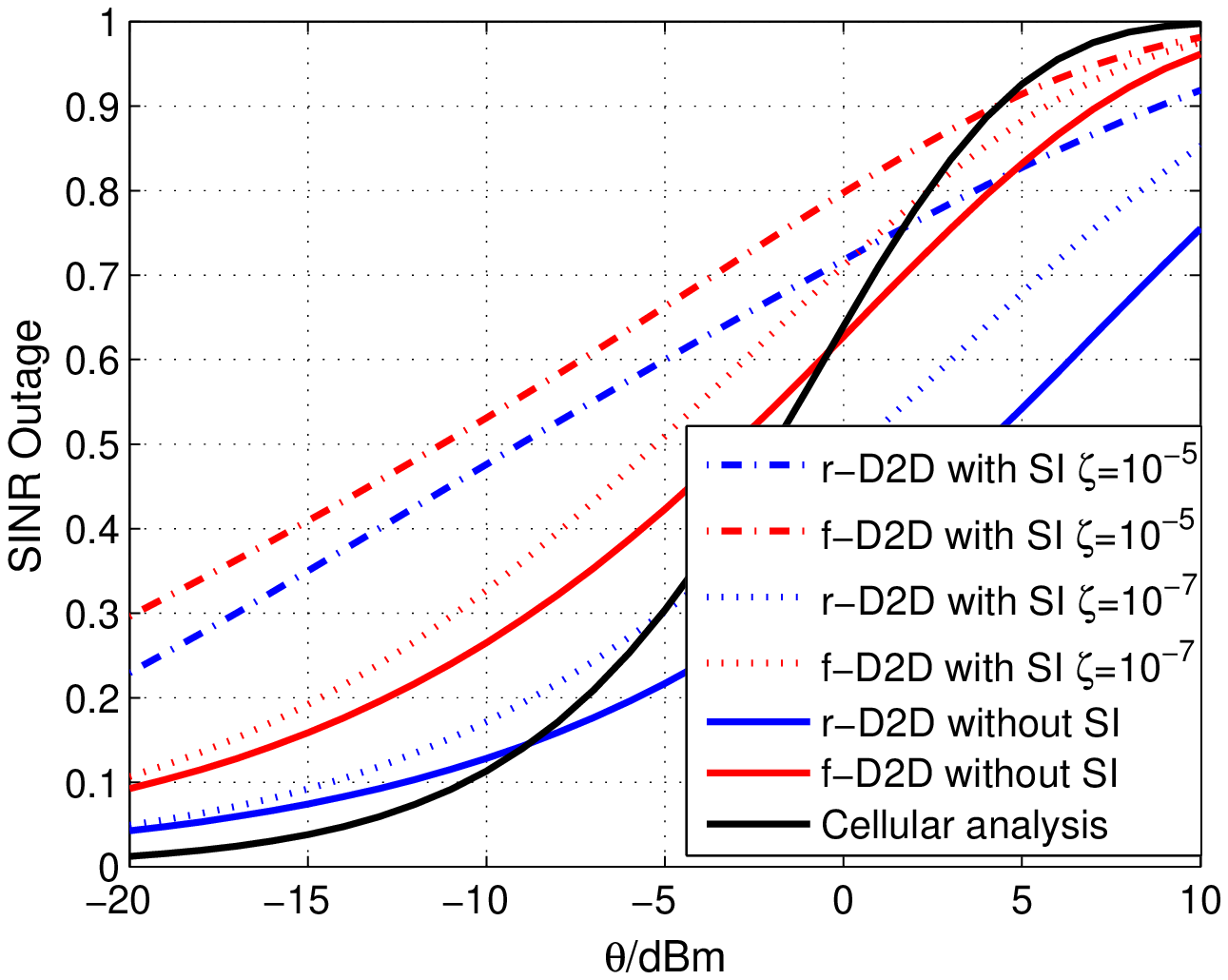}
\caption{SINR Outage vs. $\theta$ with $T_d=0.2$, $r_1=0.2$, $r_2=0.2$, and different $\zeta$.}\label{out_all_zeta}
\end{minipage}\\ 
\begin{minipage}[htb]{0.9\linewidth}
\centering\includegraphics[scale=0.5]{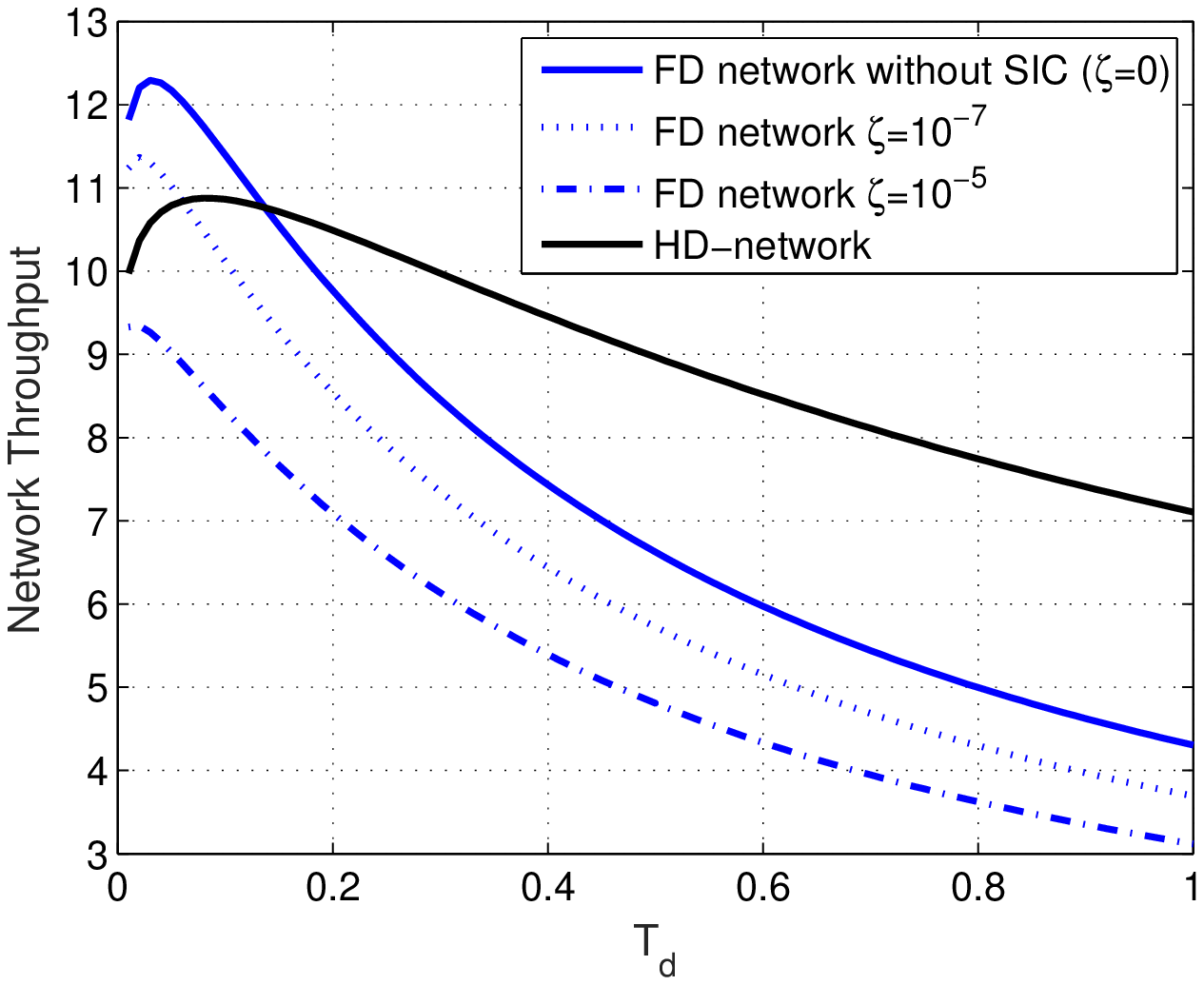}
\caption{$\mathcal{T}_n$ vs. $T_d$ with $r_1=1$, $r_2=2$, and different $\zeta$.}\label{vsTd_Tn_zeta}
\end{minipage}\\
\begin{minipage}[htb]{0.9\linewidth}
\centering\includegraphics[scale=0.5]{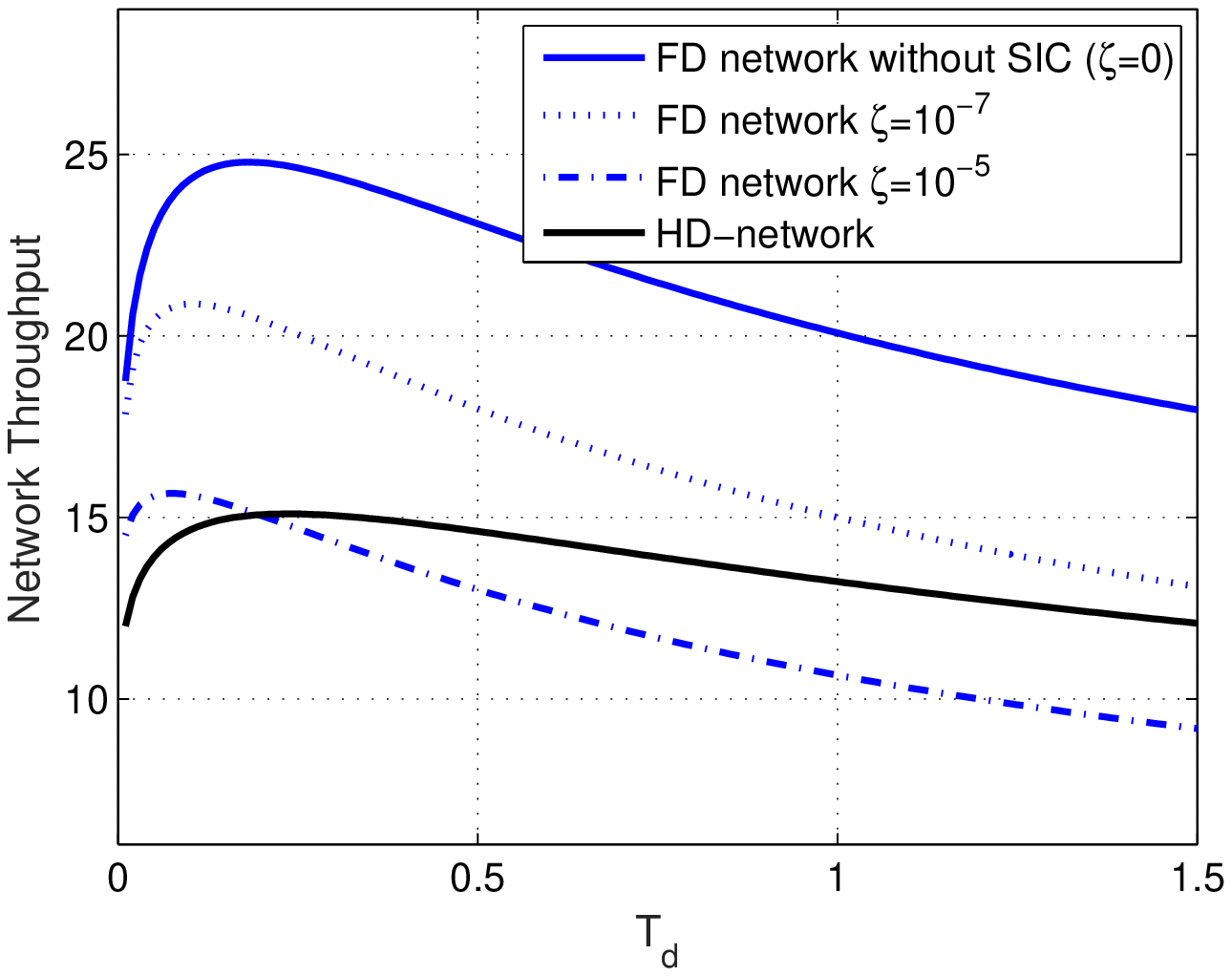}
\caption{$\mathcal{T}_n$ vs. $T_d$ with $r_1=r_2=0.2$ and different $\zeta$.}\label{vsTd_Tn_zeta_good}
\end{minipage}
\end{figure}

\subsection{{Effects of the D2D link distance distribution}}
{Finally, we inspect the effect of the link distance distribution parameter $\omega$ on the network throughput in Fig. \ref{vsTd_Tn_ww}. The figure shows that increasing $\omega$ increases $\mathcal{T}_n$ for a given $T_d$. This can be explained by the fact that larger $\omega$ values give higher weights to shorter distances, which results in less D2D transmission power due to the employed power control, and hence, less network interference and improved network throughput.}

\begin{figure}
\begin{minipage}[htb]{0.9\linewidth}
\centering\includegraphics[scale=0.5]{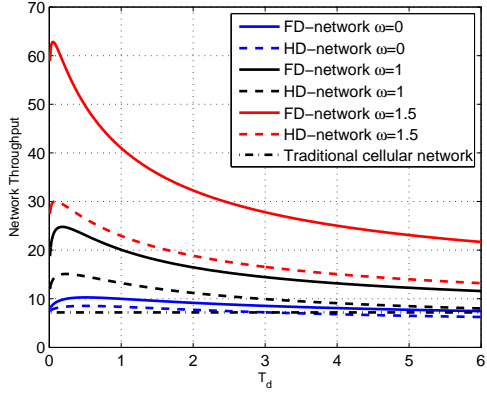}
\caption{$\mathcal{T}_n$ vs. $T_d$ with $r_1=r_2=0.2$ and different $\omega$.}\label{vsTd_Tn_ww}
\end{minipage}
\end{figure}

\section{Conclusion}\label{Conc}

{This paper presents a tractable framework for large-scale cellular networks overlaid with FD-D2D UEs that have imperfect SIC capabilities and a tunable D2D link distance distribution. We first propose a flexible network design where the flexibility comes from imposing tunable design variables that control the extent to which D2D communication is enabled in the network along with the interference protection provided for cellular users. We also propose a disjoint mode selection for the forward (f-D2D) and reverse (r-D2D) links, which depends on their relative positions from the nearest BS. To carry out our analysis, we propose an accurate approximation for the PDF of the distance between the r-D2D UE and its nearest BS. We then characterize the aggregate interference and derive the outage probability and ergodic rate. The results show that enforcing all potential D2D links to operate in D2D can severely degrade the network performance due to the imposed interference. Hence, the extent to which the D2D is enabled in the network has to be carefully tuned to balance the tradeoff between spatial frequency reuse and aggregate interference level. Due to the imposed aggregate interference, the FD-D2D communication does not double the network rate when compared to the HD-D2D operation even with perfect SIC at the optimal design variables. Nevertheless, FD-communication offers non-trivial gains compared to its HD counterpart, if the design parameters are carefully selected {({64\%} in Fig. \ref{vs_Td_Tn})}. In the case of imperfect SIC, a minimum level of SIC is required to achieve gains from employing FD-D2D compared to the HD-network. {However, if the network parameters are tuned carefully, this minimum level of SIC can be decreased.} 
Finally, we investigate the effect of the link distance distribution on the FD-network performance and show its prominent effect. While this paper shows potentials for FD-D2D communication, it also highlights the importance of sophisticated interference management to maintain an acceptable outage probability and boost the harvested FD gains.}





\appendices
\section{Proof of Proposition 1}
The exact area of the shaded crescent, given $r_c$, $r_d$ and $\bar{\theta}$, can be found by $ A_{r_c,r_d,\bar{\theta}}$, where $ A_{r_c,r_d,\bar{\theta}}= r_{c_2}^2 (\phi - \frac{\sin(2\phi)}{2}) - r_c^2 (\bar{\theta} - \frac{\sin(2 \bar{\theta})}{2})$. The angle $\phi=\pi-\arccos(\frac{r_{c_2}^2+r_d^2-r_c^2}{2r_{c_2}r_d})$ and the distribution $f_{\bar{\theta}}(\bar{\theta})=\frac{1}{\pi}, \; 0\leq \bar{\theta} \leq \pi$ is used instead of $f_{\theta}(\theta)$ due to symmetry. The average area, $A$, of the shaded crescent is found numerically as,
\begin{align*}
&A= \int_0^{\pi} \int_0^{\bar{R}} \int_0^{\infty} f_{r_c}(r_c) f_{r_d}(r_d) {f_{\bar{\theta}}(\bar{\theta})} A_{r_c,r_d,\bar{\theta}} \; dr_c dr_d d\bar{\theta}.
\end{align*}
Define $\mathcal{P}_{r_e \neq r_{c_2}}$ as the probability that at least one BS lies in the crescent in Fig. \ref{crescents}. Due to the PPP assumption the number of BSs in an area is a Poisson RV and $\mathbb{P}(\text{no BSs lie in crescent})=e^{-\lambda A}$. Therefore, by definition $\mathcal{P}_{r_e \neq r_{c_2}}=1-e^{-\lambda A}$.\\
The probability that the f-D2D UE lies closer to the r-D2D UE than its nearest BS, $\mathcal{P}_{r_c>r_d}$, is:
{\begin{align*}
&\mathcal{P}_{r_c>r_d}=\mathbb{P}(r_c>r_d)=\int_0^{\bar{R}} (1-F_{r_c}(x)) f_{r_d}(x)dx\\
&=\int_0^{\bar{R}}  \frac{(2-\omega)x^{1-\omega}}{\bar{R}^{2-\omega}} e^{-\pi \lambda x^2}  dx \\
&=\frac{2-\omega}{2\bar{R}^{2-\omega}} (\pi \lambda)^{\frac{\omega-2}{2}} [\Gamma(\frac{2-\omega}{2})-\Gamma(\frac{2-\omega}{2}, \pi \lambda \bar{R}^2)].
\end{align*}}
Using Jensen's inequality,
\begin{align*}
&\mathbb{E}[r_{c_2}]=\mathbb{E}[\sqrt{r_c^2 + r_d^2 -2 r_c r_d \cos \theta} \;] \leq \sqrt{\mathbb{E}[r_{c_2}^2]}. 
\end{align*}
Additionally, by Jensen's $\mathbb{E}[X^2] \geq (\mathbb{E}[X])^2$, and $\mathbb{E}[r_c r_d \cos \theta]=0$ due to independence. We thus approximate the mean of $r_{c_2}$ as,
{\begin{align*}
&\mu_{r_{c_2}}=\sqrt{(\mathbb{E}[r_c])^2+(\mathbb{E}[r_d])^2}=\sqrt{\frac{1}{4 \lambda} + \Big(\frac{2-\omega}{3-\omega}\bar{R}\Big)^2}.
\end{align*}}
Similarly,$\: \mu_{r_{c_2|r_c>r_d}}=\sqrt{(\mathbb{E}[r_c|r_c>r_d])^2 + (\mathbb{E}[r_d|r_c>r_d])^2}$ and
$\mu_{r_{c_2|r_c<r_d}}=\sqrt{(\mathbb{E}[r_c|r_c<r_d])^2 + (\mathbb{E}[r_d|r_c<r_d])^2}$.

The conditional PDFs are evaluated as follows,
\footnotesize
{\begin{align*}
&f_{r_d|r_c>r_d}(x)=\frac{\int_x^\infty f_{r_d|r_c}(x|y)  f_{r_c}(y) dy}{\mathbb{P}(r_d<r_c)} \\ 
&=\frac{2 x^{1-\omega} (\pi \lambda)^{\frac{2-\omega}{2}} e^{-\pi \lambda x^2} }{[\Gamma(\frac{2-\omega}{2})-\Gamma(\frac{2-\omega}{2}, \pi \lambda \bar{R}^2)]} \;\; 0\leq x \leq \bar{R}\\ 
&f_{r_c|r_c>r_d}(x)=\frac{\int_0^{\min(x,\bar{R})} f_{r_c|r_d}(x|y) f_{r_d}(y) dy}{\mathbb{P}(r_d<r_c)} \\
&= \frac{ 4 (\pi \lambda)^{\frac{4-\omega}{2}} e^{-\pi \lambda x^2} x (\min(x,\bar{R}))^{2-\omega}}{(2-\omega)[\Gamma(\frac{2-\omega}{2})-\Gamma(\frac{2-\omega}{2}, \pi \lambda \bar{R}^2)]} \;\; 0\leq x \leq \infty \\ 
&f_{r_d|r_c<r_d}(x)=\frac{\int_0^x f_{r_d|r_c}(x|y)  f_{r_c}(y) dy}{\mathbb{P}(r_d>r_c)}\\
&= \frac{2-\omega}{\bar{R}^{2-\omega}}\frac{ x^{1-\omega} (1-e^{-\pi \lambda x^2})}{(1-\mathcal{P}_{r_c>r_d})} \;\; 0\leq x \leq \bar{R} \\ 
&f_{r_c|r_c<r_d}(x)=\frac{\int_x^{\bar{R}} f_{r_c|r_d}(x|y)  f_{r_d}(y) dy}{\mathbb{P}(r_d>r_c)}\\
&=\frac{ 2 \pi \lambda x (\bar{R}^{2-\omega}-x^{2-\omega})}{  e^{\pi \lambda x^2}\bar{R}^{2-\omega}(1-\mathcal{P}_{r_c>r_d})} \;\; 0\leq x \leq \bar{R}. 
\end{align*}}
\normalsize
\noindent By integrating, the conditional expectations follow as,\\
{$\mathbb{E}[r_d|r_c>r_d]= \frac{1}{\sqrt{\pi \lambda}}\frac{\Gamma(\frac{3-\omega}{2})-\Gamma(\frac{3-\omega}{2}, \pi \lambda \bar{R}^2)}{\Gamma(\frac{2-\omega}{2})-\Gamma(\frac{2-\omega}{2}, \pi \lambda \bar{R}^2)}$} \\ 
{$\mathbb{E}[r_c|r_c>r_d]=\frac{4 (\pi \lambda)^{\frac{4-\omega}{2}}}{(2-\omega)[\Gamma(\frac{2-\omega}{2})-\Gamma(\frac{2-\omega}{2}, \pi \lambda \bar{R}^2)]} \times$ \\
$\Big[\frac{\Gamma(\frac{5-\omega}{2})-\Gamma(\frac{5-\omega}{2}, \pi \lambda \bar{R}^2)}{2 (\pi \lambda)^{\frac{5-\omega}{2}}} + \bar{R}^{2-\omega} \Big( \frac{\text{erfc}(\bar{R} \sqrt{\pi\lambda})}{4\pi \lambda^{1.5}} + \frac{\bar{R} e^{-\pi\lambda \bar{R}^2}}{2 \pi \lambda} \Big)\Big] $}\\ 
{$\mathbb{E}[r_d|r_c<r_d]= \frac{2-\omega}{(1-\mathcal{P}_{r_c>r_d})}\Big(\frac{\bar{R}}{3-\omega}-\frac{(\pi \lambda)^{\frac{\omega-3}{2}}}{2\bar{R}^{2-\omega}}\big[\Gamma(\frac{3-\omega}{2})-\Gamma(\frac{3-\omega}{2}, \pi \lambda \bar{R}^2)\big]\Big)$}\\ 
{$\mathbb{E}[r_c|r_c<r_d]=  \frac{\frac{\text{erf}(\sqrt{\pi \lambda}\bar{R})}{2\sqrt{\lambda}}-\bar{R} e^{-\pi \lambda \bar{R}^2}}{(1-\mathcal{P}_{r_c>r_d})} - \frac{[\Gamma(\frac{5-\omega}{2})-\Gamma(\frac{5-\omega}{2}, \pi \lambda \bar{R}^2)]}{(\pi \lambda)^{\frac{3-\omega}{2}} \bar{R}^{2-\omega}(1-\mathcal{P}_{r_c>r_d})}$.}  

When $r_c>r_d$ the mean of $r_e (\neq r_{c_2})$ is approximated by $\frac{\mu_{r_{c_2|r_c>r_d}} + \mathbb{E}[r_c|r_c>r_d]-\mathbb{E}[r_d|r_c>r_d]}{2}$. This is because the r-D2D UE lies outside the crescent and we find $r_e(\neq r_{c_2})$ by averaging between the inner and outer radius of the crescent due to the distribution of the BSs being homogeneous. When $r_c<r_d$, the r-D2D UE lies inside the crescent. To approximate the mean of $r_e \neq r_{c_2}$, the averaging (due to homogeneity) is done assuming the distance from the r-D2D UE to the boundary of the arc varies from 0 to $r_d-r_c$ half the time and from 0 to $r_{c_2}$ for the other half; this results in the term $\frac{\mu_{r_{c_2}|r_c<r_d} + \mathbb{E}[r_d|r_c<r_d]-\mathbb{E}[r_c|r_c<r_d]}{4}$. These result in the approximation in \eqref{mu_re_eqn}.

\section{Proof of Lemma 1 \& Lemma 2}
\subsubsection{Lemma 1} an r-D2D UE transmits if it satisfies the maximum transmit power constraint and the IP-condition. Hence,
\small
{\begin{align*}
&\mathcal{P}_e = \mathbb{P}( r_d^{\eta_d} \rho_e < P_u \; \cap \;  r_d^{\eta_d} \rho_e < T_d r_e^{\eta_c} \rho_c   )\\
&= \mathbb{P} \Big( r_d < \Big( \frac{P_u}{\rho_e}\Big)^{\frac{1}{\eta_d}}    \; \cap \; r_e > \Big( \frac{r_d^{\eta_d}\rho_e}{T_d\rho_c} \Big)^{\frac{1}{\eta_c}} \; \Big)\\
&= \int_0^{\big(\frac{P_u}{\rho_e}\big)^{\frac{1}{\eta_d}}} f_{r_d}(r) [1-F_{r_e}\Big(\Big( \frac{r^{\eta_d}\rho_e}{T_d \rho_c} \Big)^{\frac{1}{\eta_c}} \Big) ] dr.
\end{align*}}
\normalsize
\noindent Lemma 1 is obtained by evaluating the above integral with $f_{r_d}(\cdot)$ given in \eqref{power_distance} and $F_{r_e}(\cdot)$ obtained from \eqref{eqn_fre}.
\subsubsection{Lemma 2}  an f-D2D UE transmits if it satisfies the IP-condition and the maximum transmit power constraint. Hence,
\small
{\begin{align*}
&\mathcal{P}_d = \mathbb{P}(r_d^{\eta_d} \rho_d < T_d r_c^{\eta_c} \rho_c \; \cap \; r_d^{\eta_d} \rho_d < P_u)\\
&= \mathbb{P}\Big(r_c > \Big( \frac{r_d^{\eta_d}\rho_d}{T_d\rho_c} \Big)^{\frac{1}{\eta_c}} \; \cap \; r_d < \Big( \frac{P_u}{\rho_d}\Big)^{\frac{1}{\eta_d}} \; \Big).
\end{align*}}
\normalsize
\noindent This can be calculated using similar steps to the proof of the r-D2D case shown above.

\section{Proof of Lemma 3}
{Denote by $Z$ the maximum allowed distance between a D2D pair for the f-D2D UE to be transmitting i.e. $Z=\min\big(\big(\frac{P_u}{\rho_d}\big)^{\frac{1}{\eta_d}}, \big(\frac{T_d \rho_c r_c^{\eta_c}}{\rho_d} \big)^{\frac{1}{\eta_d}}\big)$ and $f_Z(z)=2 \pi \lambda \frac{\eta_d}{\eta_c} (\frac{\rho_d}{T_d \rho_c})^{\frac{2}{\eta_c}}\frac{ z^{\frac{2 \eta_d}{\eta_c}-1}  e^{-\pi \lambda (\frac{z^{\eta_d}\rho_d}{T_d \rho_c})^{\frac{2}{\eta_c}} }}{1- \dot{q}} \; 0\leq z \leq \big(\frac{P_u}{\rho_d}\big)^{\frac{1}{\eta_d}}$, where $\dot{q}=e^{-\pi \lambda (\frac{P_u}{\rho_c T_d})^{\frac{2}{\eta_c}} }$. For a D2D pair to be operating in FD we require both the r-D2D and f-D2D UE to be transmitting. Thus, 
\small
\begin{align*}
&\mathcal{P}_{FD} = \mathbb{P}(\text{r-D2D is transmitting} \cap \text{f-D2D is transmitting})\\
&= \underbrace{\mathbb{P}\Big(r_d^{\eta_d}  < \frac{P_u}{\rho_e}  \cap   r_d^{\eta_d} < \frac{T_d r_e^{\eta_c} \rho_c}{\rho_e } \Big| r_d<Z  \Big)}_{\mathcal{P}_1} \mathcal{P}_d,
\end{align*}
\normalsize
where $\mathcal{P}_1$ denotes the conditional probability $\mathbb{P}(\text{r-D2D is transmitting} | \text{f-D2D is transmitting})$ and $\mathcal{P}_d$ is given in Lemma 2. We evaluate $\mathcal{P}_1$ as,
\footnotesize
\begin{align*}
&\int_0^{\infty} f_{r_e}(g) \int_0^{{\big(\frac{\min(P_u,T_d g^{\eta_c}\rho_c)}{\rho_e}\big)}^{\frac{1}{\eta_d}}} \int_x^{(\frac{P_u}{\rho_d})^{\frac{1}{\eta_d}}} \frac{ f_{r_d|z}(x|z) f_z(z) dz}{\mathcal{P}_d} dx\; dg\\
&=\int_0^{\infty} \frac{f_{r_e}(g)/(1- \dot{q})  }{ (\bar{R}^{2-\omega} \mathcal{P}_d)} \Bigg( \gamma \Big(\frac{(2-\omega)\eta_c}{2\eta_d}, \pi \lambda \Big(\frac{ \min(P_u, T_d g^{\eta_c} \rho_c)}{T_d \rho_c \rho_e/\rho_d}\Big)^{\frac{2}{\eta_c}}\Big)  \times \\ 
& \frac{(2-\omega)\eta_c}{2\eta_d} \Big(\frac{T_d \rho_c}{\rho_d (\pi \lambda)^{\frac{\eta_c}{2}}}\Big)^{\frac{2-\omega}{\eta_d}}  - \dot{q}\times\Big(\frac{\min(P_u, T_d g^{\eta_c} \rho_c)}{\rho_e} \Big)^{\frac{2-\omega}{\eta_d}} \Bigg)  dg.
\end{align*} }
\normalsize
\section{Proof of Lemma 4}
Denote by $X_c$ and $X_d$ the unconditional transmit powers required to invert the channel to the nearest BS and r-D2D UE, respectively; hence, $X_c=r_c^{\eta_c}\rho_c$ and $X_d=r_d^{\eta_d}\rho_d$. Using the PDFs of $r_c$ and $r_d$ we obtain $f_{X_c}(x)=\frac{2 \pi \lambda x^{\frac{2}{\eta_c}-1}}{\eta_c \rho_c^{\frac{2}{\eta_c}}} e^{-\pi \lambda (\frac{x}{\rho_c})^{\frac{2}{\eta_c}}} \; 0\leq x \leq \infty$ and {$f_{X_d}(x)=\frac{2-\omega}{ \bar{R}^{2-\omega}}\frac{x^{\frac{2-\omega}{\eta_d}-1}}{\eta_d \rho_d^{\frac{2-\omega}{\eta_d}}}\; 0\leq x \leq {\bar{R}}^{\eta_d} \rho_d$}. As successful communication requires satisfying the IP-condition and the maximum transmit power constraint, the PDF of the transmit power in the f-D2D mode is given by,
{\begin{align*}
&f_{P_d}(x) = \frac{\int_x^\infty f_{X_d|T_d X_c}(x|y)  f_{X_c}(y) dy}{\mathbb{P}(X_d\leq T_d X_c)} \\
&= \frac{2-\omega}{ \bar{R}^{2-\omega}}\frac{x^{\frac{2-\omega}{\eta_d}-1}}{\eta_d \rho_d^{\frac{2-\omega}{\eta_d}} \mathcal{P}_d} \int_{\frac{x}{T_d}}^{\infty} \frac{2 \pi \lambda y^{\frac{2}{\eta_c}-1}}{\eta_c \rho_c^{\frac{2}{\eta_c}}} e^{-\pi \lambda (\frac{y}{\rho_c})^{\frac{2}{\eta_c}}} dy\\
&= \frac{2 x^{\frac{2-\omega}{\eta_d} - 1} e^{-\pi \lambda {(\frac{x}{T_d\rho_c})}^{\frac{2}{\eta_c}}} {(\pi \lambda)}^{\frac{(2-\omega)\eta_c}{2\eta_d}}}{\eta_c {(\rho_c T_d)}^{\frac{2-\omega}{\eta_d}} \gamma\Big(\frac{(2-\omega)\eta_c}{2\eta_d},\pi \lambda \Big({\frac{P_u}{\rho_c T_d}}\Big)^{\frac{2}{\eta_c}}\Big)} , \;\; 0\leq x \leq P_u. 
\end{align*} }
We obtain $\mathbb{E}[P_d^{\alpha}]$ as $\int_0^{P_u} x^{\alpha} f_{P_d}(x)dx$.

\section{Proof of Lemma 5}
Denote by $X_i$ and $X_e$ the unconditional transmit powers required to invert the channel to the nearest BS and f-D2D UE, respectively; hence, $X_i=r_e^{\eta_c}\rho_c$ and $X_e=r_d^{\eta_d}\rho_e$. Using the PDFs of $r_e$ and $r_d$ we obtain $f_{X_i}(x)=\frac{2 b x^{\frac{2}{\eta_c}-1}}{\eta_c \rho_c^{\frac{2}{\eta_c}}} e^{-b (\frac{x}{\rho_c})^{\frac{2}{\eta_c}}} \;\; 0\leq x \leq \infty$, and {$f_{X_e}(x)=\frac{2-\omega}{ \bar{R}^{2-\omega}}\frac{x^{\frac{2-\omega}{\eta_d}-1}}{\eta_d \rho_e^{\frac{2-\omega}{\eta_d}}} \; 0\leq x \leq \bar{R}^{\eta_d} \rho_e$}. As successful communication in the r-D2D mode requires satisfying the IP-condition and the maximum transmit power constraint, the PDF of the transmit power in the r-D2D mode is,
{\begin{align*}
&f_{P_e}(x) = \frac{\int_x^\infty f_{X_e|T_d X_i}(x|y)  f_{X_i}(y) dy}{\mathbb{P}(X_e\leq T_d X_i)} \\
&= \frac{2-\omega}{ \bar{R}^{2-\omega}}\frac{x^{\frac{2-\omega}{\eta_d}-1}}{\eta_d \rho_e^{\frac{2-\omega}{\eta_d}}\mathcal{P}_e} \int_{\frac{x}{T_d}}^{\infty} \frac{2 b y^{\frac{2}{\eta_c}-1}}{\eta_c \rho_c^{\frac{2}{\eta_c}}} e^{-b (\frac{y}{\rho_c})^{\frac{2}{\eta_c}}} dy\\
&=  \frac{2 (T_d\rho_c)^{\frac{(\omega-2)}{\eta_d}} b^{\frac{(2-\omega)\eta_c}{2 \eta_d}} x^{\frac{(2-\omega)}{\eta_d}-1} e^{-b(\frac{x}{T_d\rho_c})^{\frac{2}{\eta_c}}} }{\eta_c  \gamma \Big(\frac{(2-\omega)\eta_c}{2\eta_d}, b \Big(\frac{P_u}{T_d\rho_c}\Big)^{\frac{2}{\eta_c}} \Big)}, \;\;\; 0\leq x \leq P_u.
\end{align*} }
We obtain $\mathbb{E}[P_e^{\alpha}]$ by $\int_0^{P_u} x^{\alpha} f_{P_e}(x)dx$.
\section{Proof of Lemma 6}
Due to the PPP assumption, the cellular link distance $r_c$ follows a Rayleigh distribution mentioned in Section \ref{SysMod}. Denote by $X_c=r_c^{\eta_c} \rho_c$ the unconditional transmit power required to invert the channel to the nearest BS, where $f_{X_c}(x)=\frac{2 \pi \lambda x^{\frac{2}{\eta_c}-1}}{\eta_c \rho_c^{\frac{2}{\eta_c}}} e^{-\pi \lambda (\frac{x}{\rho_c})^{\frac{2}{\eta_c}}} \;\; 0\leq x \leq \infty$. Since successful communication requires satisfying the maximum transmit power constraint, we require $P_c<P_u$. Hence, the PDF of the transmit power is obtained as 
\begin{align*}
f_{P_c}(x) &= \frac{f_{X_c}(x)}{  \int_0^{P_u} f_{X_c}(y) dy }= \frac{2 \pi \lambda x^{\frac{2}{\eta_c} - 1} e^{-\pi \lambda {(\frac{x}{\rho_c})}^{\frac{2}{\eta_c}}}}{\eta_c \rho_c^{\frac{2}{\eta_c}} \Big(1-e^{-\pi \lambda{(\frac{P_u}{\rho_c})}^{\frac{2}{\eta_c}}}\Big)}, 
\end{align*}
where, $0 \leq x \leq P_u$. We obtain $\mathbb{E}[P_c^{\alpha}]$ by $\int_0^{P_u} x^{\alpha} f_{P_c}(x)dx$.

\section{Proof of Lemma 7}\label{proof7}
The LTs of the interferences experienced by receivers in mode $\chi$ from UEs in mode $\kappa$ are evaluated as,
\small
\begin{align*}
& \mathcal{L}_{\mathcal{I}_{\kappa \chi}}(s)=\mathbb{E}_{\widetilde{\Phi}_{\kappa},P_{\kappa},h} [e^{-s \sum_{u_j \in \widetilde{\Phi}_{\kappa}} P_{\kappa_j} h_j u_j^{-\eta_{\chi}}}]\\
&\stackrel{(1)}{=} \mathbb{E}_{\widetilde{\Phi}_{\kappa}}[\mathbb{E}_{P_{\kappa},h} [e^{-s \sum_{u_j \in \widetilde{\Phi}_{\kappa}} P_{{\kappa}_j} h_j u_j^{-\eta_{\chi}}}]]\\
&= \mathbb{E}_{\widetilde{\Phi}_{\kappa}}[\prod_{u_j \in \widetilde{\Phi}_{\kappa}} \mathbb{E}_{P_{\kappa},h} [e^{-s  P_{{\kappa}_j} h_j u_j^{-\eta_{\chi}}}]]\\
&\stackrel{(2)}{=} \exp \Big( -2 \pi \mathcal{U}_{\kappa} \int_{\text{IP-boundary}_{\kappa \chi}}^{\infty} \mathbb{E}_{P_{\kappa},h} [1-e^{-s  P_{\kappa} h u^{-\eta_{\chi}}}] u \; du \Big)\\
&\stackrel{(3)}{=} \exp \Big( -2 \pi \mathcal{U}_{\kappa} \int_{\text{IP-boundary}_{\kappa \chi}}^{\infty} \mathbb{E}_{P_{\kappa}} [1-\frac{1}{1+s  P_{\kappa} u^{-\eta_{\chi}}}] u \; du \Big),
\end{align*}
\normalsize
where $\text{IP-boundary}_{\kappa \chi}=0$ for $\{ \chi \in \{d,e \}, \forall \kappa \}$ as the f-D2D and r-D2D modes do not offer any IP to their receivers. For $ \kappa \in \{d,e\}$, $\text{IP-boundary}_{\kappa c}=\Big(\frac{P_{\kappa}}{\rho_c T_d}\Big)^{\frac{1}{\eta_c}}$ as the BSs have IP from the transmitting f-D2D and r-D2D UEs. Additionally, an IP to BSs from cellular interferers exists as only one cellular UE serves the BS on a channel at a time; thus, interfering cellular UEs have $\text{IP-boundary}_{cc}=\Big(\frac{P_{c}}{\rho_c}\Big)^{\frac{1}{\eta_c}}$. The intensity of the interferers in mode $\kappa$ is denoted by $\mathcal{U}_{\kappa}$, hence $\mathcal{U}_{\kappa}$ for $\kappa \in \{d,e\}$ is $\mathcal{U}_d$ and $\mathcal{U}_e$, respectively, while $\mathcal{U}_c=\lambda$. Due to the independence of fading, we have (1), and (2) follows from using the PGFL of the PPP. Using the MGF of $h \sim \exp(1)$, (3) is obtained. Continuing the integration in (3) and applying a change of variables, we arrive at the LTs of the interferences in Lemma 7.

\bibliographystyle{IEEEtran}
\bibliography{refsFD}

\begin{thebibliography}{10}
\providecommand{\url}[1]{#1}
\csname url@samestyle\endcsname
\providecommand{\newblock}{\relax}
\providecommand{\bibinfo}[2]{#2}
\providecommand{\BIBentrySTDinterwordspacing}{\spaceskip=0pt\relax}
\providecommand{\BIBentryALTinterwordstretchfactor}{4}
\providecommand{\BIBentryALTinterwordspacing}{\spaceskip=\fontdimen2\font plus
\BIBentryALTinterwordstretchfactor\fontdimen3\font minus
  \fontdimen4\font\relax}
\providecommand{\BIBforeignlanguage}[2]{{%
\expandafter\ifx\csname l@#1\endcsname\relax
\typeout{** WARNING: IEEEtran.bst: No hyphenation pattern has been}%
\typeout{** loaded for the language `#1'. Using the pattern for}%
\typeout{** the default language instead.}%
\else
\language=\csname l@#1\endcsname
\fi
#2}}
\providecommand{\BIBdecl}{\relax}
\BIBdecl

\bibitem{3F}
X.~Lin, J.~Andrews, and A.~Ghosh, ``Spectrum sharing for device-to-device
  communication in cellular networks,'' \emph{IEEE Trans. Wireless Commun.},
  vol.~13, no.~12, pp. 6727--6740, Dec. 2014.

\bibitem{3G2}
N.~Lee, X.~Lin, J.~Andrews, and R.~Heath, ``Power control for {D2D} underlaid
  cellular networks: Modeling, algorithms, and analysis,'' \emph{IEEE J.
  Select. Areas Commun.}, vol.~33, no.~1, pp. 1--13, Jan. 2015.

\bibitem{3H}
C.-H. Yu, O.~Tirkkonen, K.~Doppler, and C.~Ribeiro, ``On the performance of
  device-to-device underlay communication with simple power control,'' in
  \emph{Proc. of IEEE 69th Vehicular Technology Conference (VTC Spring 2009)},
  Apr. 2009, pp. 1--5.

\bibitem{3I}
Z.-S. Syu and C.-H. Lee, ``Spatial constraints of device-to-device
  communications,'' in \emph{First International Black Sea Conference on
  Communications and Networking (BlackSeaCom)}, Jul. 2013, pp. 94--98.

\bibitem{3J}
Z.~Liu, T.~Peng, Q.~Lu, and W.~Wang, ``Transmission capacity of {D2D}
  communication under heterogeneous networks with dual bands,'' in \emph{7th
  International ICST Conference on Cognitive Radio Oriented Wireless Networks
  and Communications (CROWNCOM)}, Jun. 2012, pp. 169--174.

\bibitem{3Kjrnl}
\BIBentryALTinterwordspacing
J.~Guo, S.~Durrani, X.~Zhou, and H.~Yanikomeroglu, ``Device-to-device
  communication underlaying a finite cellular network region,'' \emph{{IEEE}
  Trans. Wireless Commun., submitted}, 2015. [Online]. Available:
  \url{http://arxiv.org/abs/1510.03162}
\BIBentrySTDinterwordspacing

\bibitem{d2dSurvey1}
J.~Liu, N.~Kato, J.~Ma, and N.~Kadowaki, ``Device-to-device communication in
  {LTE}-advanced networks: A survey,'' \emph{IEEE Commun. Surveys and
  Tutorials}, vol.~PP, no.~99, pp. 1--1, 2014.

\bibitem{d2dSurvey2}
A.~Asadi, Q.~Wang, and V.~Mancuso, ``A survey on device-to-device communication
  in cellular networks,'' \emph{IEEE Commun. Surveys and Tutorials}, vol.~16,
  no.~4, pp. 1801--1819, 2014.

\bibitem{d2dSurvey3}
P.~Mach, Z.~Becvar, and T.~Vanek, ``In-band device-to-device communication in
  {OFDMA} cellular networks: A survey and challenges,'' \emph{IEEE Commun.
  Surveys and Tutorials}, vol.~PP, no.~99, 2015.

\bibitem{d2dSurvey4}
R.~Alkurd, R.~Shubair, and I.~Abualhaol, ``Survey on device-to-device
  communications: Challenges and design issues,'' in \emph{Proc. of IEEE 12th
  Int. New Circuits Syst. Conf. (NEWCAS)}, Jun. 2014, pp. 361--364.

\bibitem{4A}
M.~Belleschi, G.~Fodor, and A.~Abrardo, ``Performance analysis of a distributed
  resource allocation scheme for {D2D} communications,'' in \emph{Proc. of IEEE
  Global Communications Conference Workshops (GC Wkshps)}, Dec. 2011, pp.
  358--362.

\bibitem{5G_Jeff}
J.~Andrews, S.~Buzzi, W.~Choi, S.~Hanly, A.~Lozano, A.~Soong, and J.~Zhang,
  ``What will 5{G} be?'' \emph{IEEE J. Select. Areas Commun.}, vol.~32, no.~6,
  pp. 1065--1082, Jun. 2014.

\bibitem{5G_Mag1}
H.~Elsawy, H.~Dahrouj, T.~Al-Naffouri, and M.-S. Alouini, ``Virtualized
  cognitive network architecture for 5{G} cellular networks,'' \emph{IEEE
  Commun. Mag.}, vol.~53, no.~7, pp. 78--85, Jul. 2015.

\bibitem{1sic1_fd5}
\BIBentryALTinterwordspacing
J.~I. Choi, M.~Jain, K.~Srinivasan, P.~Levis, and S.~Katti, ``Achieving single
  channel, full duplex wireless communication,'' in \emph{Proc. of the
  Sixteenth Annual International Conference on Mobile Computing and Networking
  (MobiCom10)}, 2010, pp. 1--12. [Online]. Available:
  \url{http://doi.acm.org/10.1145/1859995.1859997}
\BIBentrySTDinterwordspacing

\bibitem{2sic2_fd5}
M.~Duarte and A.~Sabharwal, ``Full-duplex wireless communications using
  off-the-shelf radios: Feasibility and first results,'' in \emph{the Forty
  Fourth Asilomar Conference on Signals, Systems and Computers (ASILOMAR10)},
  Nov. 2010, pp. 1558--1562.

\bibitem{FD1}
S.~Kim and W.~Stark, ``Full duplex device to device communication in cellular
  networks,'' in \emph{International Conference on Computing, Networking and
  Communications (ICNC14)}, Feb. 2014, pp. 721--725.

\bibitem{FD2}
K.~Hemachandra, N.~Rajatheva, and M.~Latva-aho, ``Sum-rate analysis for
  full-duplex underlay device-to-device networks,'' in \emph{IEEE Wireless
  Communications and Networking Conference (WCNC14)}, Apr. 2014, pp. 514--519.

\bibitem{FD3}
S.~Ali, N.~Rajatheva, and M.~Latva-aho, ``Full duplex device-to-device
  communication in cellular networks,'' in \emph{European Conference on
  Networks and Communications (EuCNC14)}, Jun. 2014, pp. 1--5.

\bibitem{d2dOrig}
H.~ElSawy, E.~Hossain, and M.-S. Alouini, ``Analytical modeling of mode
  selection and power control for underlay {D2D} communication in cellular
  networks,,'' \emph{IEEE Trans. Commun.}, vol.~62, no.~11, pp. 4147--4161,
  Nov. 2014.

\bibitem{d2dICC}
K.~Ali, H.~ElSawy, and M.-S. Alouini, ``On mode selection and power control for
  uplink {D2D} communication in cellular networks,'' in \emph{Proc. of IEEE
  International Conference on Communications Workshops (ICC15)}, Jun. 2015.

\bibitem{3B1}
J.~Andrews, F.~Baccelli, and R.~Ganti, ``A tractable approach to coverage and
  rate in cellular networks,'' \emph{IEEE Trans. Commun.}, vol.~59, no.~11, pp.
  3122--3134, Nov. 2011.

\bibitem{3B2}
H.~ElSawy, E.~Hossain, and M.~Haenggi, ``Stochastic geometry for modeling,
  analysis, and design of multi-tier and cognitive cellular wireless networks:
  A survey,'' \emph{IEEE Commun. Surveys and Tutorials}, vol.~15, no.~3, pp.
  996--1019, Jul. 2013.

\bibitem{h_tut}
\BIBentryALTinterwordspacing
H.~ElSawy, A.~K. Sultan{-}Salem, M.-S. Alouini, and M.~Z. Win, ``Modeling and
  analysis of cellular networks using stochastic geometry: {A} tutorial,''
  \emph{CoRR}, vol. abs/1604.03689, 2016. [Online]. Available:
  \url{http://arxiv.org/abs/1604.03689}
\BIBentrySTDinterwordspacing

\bibitem{3B3}
A.~Guo and M.~Haenggi, ``Spatial stochastic models and metrics for the
  structure of base stations in cellular networks,'' \emph{IEEE Trans. Wireless
  Commun.}, vol.~12, no.~11, pp. 5800--5812, Nov. 2013.

\bibitem{3B5}
H.~Dhillon, R.~Ganti, F.~Baccelli, and J.~Andrews, ``Modeling and analysis of
  {K}-tier downlink heterogeneous cellular networks,'' \emph{IEEE J. Select.
  Areas Commun.}, vol.~30, no.~3, pp. 550--560, Apr. 2012.

\bibitem{3B6}
B.~Blaszczyszyn, M.~Karray, and H.~Keeler, ``Using {P}oisson processes to model
  lattice cellular networks,'' in \emph{Proc. 32nd Annual IEEE International
  Conference on Computer Communications (INFOCOM’13)}, Apr. 2013, pp.
  773--781.

\bibitem{di_renzo}
\BIBentryALTinterwordspacing
W.~Lu and M.~D. Renzo, ``Stochastic geometry modeling of cellular networks:
  Analysis, simulation and experimental validation,'' \emph{CoRR}, vol.
  abs/1506.03857, 2015. [Online]. Available:
  \url{http://arxiv.org/abs/1506.03857}
\BIBentrySTDinterwordspacing

\bibitem{FD4}
S.~Goyal, P.~Liu, S.~Hua, and S.~Panwar, ``Analyzing a full-duplex cellular
  system,'' in \emph{47th Annual Conference on Information Sciences and Systems
  (CISS13)}, Mar. 2013, pp. 1--6.

\bibitem{FD5_2}
Z.~Tong and M.~Haenggi, ``Throughput analysis for full-duplex wireless networks
  with imperfect self-interference cancellation,'' \emph{IEEE Trans. Commun.},
  vol.~63, no.~11, pp. 4490--4500, Nov 2015.

\bibitem{FD6}
J.~Lee and T.~Quek, ``Hybrid full-/half-duplex system analysis in heterogeneous
  wireless networks,'' \emph{IEEE Trans. Wireless Commun.}, vol.~14, no.~5, pp.
  2883--2895, May 2015.

\bibitem{AlAmmouri}
\BIBentryALTinterwordspacing
A.~AlAmmouri, H.~ElSawy, O.~Amin, and M.-S. Alouini, ``In-band
  {$\alpha$}-duplex scheme for cellular networks: A stochastic geometry
  approach,'' \emph{{IEEE} Trans. Wireless Commun., submitted}, 2015. [Online].
  Available: \url{http://arxiv.org/abs/1509.00976}
\BIBentrySTDinterwordspacing

\bibitem{Tsiky}
{I. Randrianantenaina}, {H. Elsawy}, and {M.-S. Alouini}, ``Limits on the
  capacity of in-band full duplex communication in uplink cellular networks,''
  in \emph{Workshop. of IEEE Global Communications Conference (Globecom'15)},
  San Diego, California, Dec. 2015.

\bibitem{d2d_dist}
J.~Lee and T.~Q.~S. Quek, ``Device-to-device communication in wireless mobile
  social networks,'' in \emph{Proc. of IEEE 79th Vehicular Technology
  Conference (VTC Spring 2014)}, May 2014, pp. 1--5.

\bibitem{3B7}
H.~ElSawy and E.~Hossain, ``On stochastic geometry modeling of cellular uplink
  transmission with truncated channel inversion power control,'' \emph{IEEE
  Trans. Wireless Commun.}, vol.~13, no.~8, pp. 4454--4469, Aug. 2014.

\bibitem{JA1}
\BIBentryALTinterwordspacing
S.~Singh, X.~Zhang, and J.~G. Andrews, ``Joint rate and {SINR} coverage
  analysis for decoupled uplink-downlink biased cell associations in hetnets,''
  \emph{CoRR}, vol. abs/1412.1898, 2014. [Online]. Available:
  \url{http://arxiv.org/abs/1412.1898}
\BIBentrySTDinterwordspacing

\end{thebibliography}

\end{document}